\documentclass[]{emulateapj}  

\usepackage{amsmath}
\usepackage{amssymb}
\usepackage{natbib}
\usepackage{apjfonts}
\usepackage{graphicx}
\usepackage{multirow}
\usepackage{booktabs}
\usepackage{bm}
\usepackage{xspace}
\usepackage{afterpage}

\usepackage{color}
\usepackage[usenames,dvipsnames]{xcolor}
\definecolor{midgray}{gray}{0.4}
\usepackage[colorlinks=true,citecolor=blue,linkcolor=magenta]{hyperref}

\newcommand{\HST}{\textit{HST}\xspace}
\newcommand{\spitzer}{\textit{Spitzer}\xspace}
\newcommand{\photz}{$z_{\mathrm{phot}}$}
\newcommand{\tphot}{\textsc{T-PHOT}\xspace}
\newcommand{\zbest}{z_{\mathrm{best}}}
\newcommand{\mubest}{\mu_{\mathrm{best}}}
\newcommand{\sfrunit}{M_{\odot}\,\mathrm{yr}^{-1}}
\newcommand{\mstar}{M^*_{\mathrm{UV}}\xspace}

\begin{document}

\title{\emph{Spitzer} UltRa Faint SUrvey Program (SURFS UP). II. IRAC-Detected Lyman-Break Galaxies at $6 \lesssim z \lesssim 10$ Behind Strong-Lensing Clusters}

\author{Kuang-Han Huang\altaffilmark{1}}
\author{Maru\v{s}a Brada\v{c}\altaffilmark{1}}
\author{Brian C. Lemaux\altaffilmark{3}}
\author{R.~E.~Ryan,~Jr.\altaffilmark{2}}
\author{Austin Hoag\altaffilmark{1}}
\author{Marco Castellano\altaffilmark{4}}
\author{Ricardo Amor\'{i}n\altaffilmark{4}}
\author{Adriano Fontana\altaffilmark{4}}
\author{Gabriel B. Brammer\altaffilmark{2}}
\author{Benjamin Cain\altaffilmark{1}}
\author{L.~M.~Lubin\altaffilmark{1}}
\author{Emiliano Merlin\altaffilmark{4}}
\author{Kasper B. Schmidt\altaffilmark{5}}
\author{Tim Schrabback\altaffilmark{6}}
\author{Tommaso Treu\altaffilmark{7}}
\author{Anthony H. Gonzalez\altaffilmark{8}}
\author{Anja von der Linden\altaffilmark{9,10,11}}
\author{Robert I. Knight\altaffilmark{1}}

\altaffiltext{1}{University of California Davis, 1 Shields Avenue, Davis, CA 95616, USA; khhuang@ucdavis.edu}
\altaffiltext{3}{Space Telescope Science Institute, 3700 San Martin Drive, Baltimore, MD 21218, USA}
\altaffiltext{2}{Aix Marseille Universit\'{e}, CNRS, LAM (Laboratoire d'Astrophysique de Marseille) UMR 7326, F-13388 Marseille, France}
\altaffiltext{4}{INAF--Osservatorio Astronomico di Roma Via Frascati 33, I-00040 Monte Porzio Catone, Italy}
\altaffiltext{5}{Department of Physics, University of California, Santa Barbara, CA 93106, USA}
\altaffiltext{6}{Argelander-Institut f\"{u}r Astronomie, Auf Dem H\"ugel 71, D-53121 Bonn, Germany}
\altaffiltext{7}{Department of Physics and Astronomy, UCLA, Los Angeles, CA 90095, USA}
\altaffiltext{8}{Department of Astronomy, University of Florida, 211 Bryant Space Science Center, Gainesville, FL 32611, USA}
\altaffiltext{9}{Department of Physics, Stanford University, 382 Via Pueblo Mall, Stanford, CA 94305, USA}
\altaffiltext{10}{Dark Cosmology Centre, Niels Bohr Institute, University of Copenhagen, Juliane Maries Vej 30, DK-2100 Copenhagen \O, Denmark}
\altaffiltext{11}{Kavli Institute for Particle Astrophysics and Cosmology, Stanford University, 382 Via Pueblo Mall, Stanford, CA 94305-4060, USA}
\email[E-mail:~]{astrokuang@gmail.com}
\begin{abstract}

We study the stellar population properties of the IRAC-detected $6 \lesssim z \lesssim 10$ galaxy candidates from the \spitzer UltRa Faint SUrvey Program (SURFS UP). Using the Lyman Break selection technique, we find a total of 17 galaxy candidates at $6 \lesssim z \lesssim 10$ from \HST images (including the full-depth images from the Hubble Frontier Fields program for MACS1149 and MACS0717) that have detections at $S/N \geq 3$ in at least one of the IRAC $3.6\mu$m and $4.5\mu$m channels. According to the best mass models available for the surveyed galaxy clusters, these IRAC-detected galaxy candidates are magnified by factors of $\sim 1.2$--$5.5$. Due to the magnification of the foreground galaxy clusters, the rest-frame UV absolute magnitudes $M_{1600}$ are between $-21.2$ and $-18.9$ mag, while their \emph{intrinsic} stellar masses are between $2 \times 10^8\,M_\odot$ and $2.9 \times 10^9\,M_\odot$. We identify two Ly$\alpha$ emitters in our sample from the Keck DEIMOS spectra, one at $z_{\text{Ly}\alpha}=6.76$ (in RXJ1347) and one at $z_{\text{Ly}\alpha}=6.32$ (in MACS0454). We find that 4 out of 17 $z \gtrsim 6$ galaxy candidates are favored by $z \lesssim 1$ solutions when IRAC fluxes are included in photometric redshift fitting. We also show that IRAC $[3.6]-[4.5]$ color, when combined with photometric redshift, can be used to identify galaxies likely with strong nebular emission lines or have obscured AGN contributions within certain redshift windows.

\end{abstract}

\section{Introduction}


Galaxies at $6 \lesssim z \lesssim 10$ are one of the frontiers in observational astronomy because they are a key player in the reionization process. It is widely postulated that galaxies provided the bulk of ionization photons, but low-level AGN activity (with their likely very high escape fractions of ionizing photons) is still a possibility (e.g., \citealt{Giallongo:2015hx}). To improve our knowledge about the importance of galaxies on reionization, we should measure their ionizing photon production rate (through their star formation rate density) and their ionizing photon escape fraction (\citealt{Robertson:2010ec}). We should also measure their stellar mass and, under reasonable assumptions about their star formation history, infer how many ionizing photons they produced in the past.


Rest-frame optical stellar emission from $z \gtrsim 6$ galaxies is crucial for stellar mass measurement; the rest-frame 4000\AA\ break shifts to $\gtrsim 3\mu m$ in the observed frame and requires deep \emph{Spitzer} observations at the moment. Over a thousand $z \gtrsim 6$ galaxy candidates have been identified in deep \HST extragalactic blank fields like CANDELS (\citealt{Koekemoer:2011br, Grogin:2011hx}), HUDF/XDF (\citealt{Beckwith:2006hp, Koekemoer:2013db, Illingworth:2013dk, Bouwens:2015gm}), BoRG (\citealt{Trenti:2011ft,Trenti:2012hg,Bradley:2012je}), and HIPPIES (\citealt{Yan:2011cz}). Among all the $z \gtrsim 6$ galaxy candidates, more than 100 of them have individual \emph{Spitzer}/IRAC detections (\citealt{Eyles:2007el,Yan:2006gy,Stark:2009eu,Labbe:2010ho,Capak:2011gd,Labbe:2013kx,RobertsBorsani:2015um,Labbe:2015tq}); their IRAC fluxes enable more robust constraints on their stellar masses. The inferred stellar masses of these $z \gtrsim 6$ galaxy candidates range from $\sim 10^9$ to $\sim 10^{11}\,M_{\odot}$, surprisingly large for a universe younger than 1 Gyr old. It is likely that the majority of IRAC-detected $z \gtrsim 6$ galaxies are at the high-mass end of the stellar mass function, although some of these galaxies likely have their IRAC fluxes boosted by strong nebular emission lines like [OIII], H$\alpha$, and H$\beta$ (e.g., \citealt{Finkelstein:2013fx, Smit:2014cg, DeBarros:2014fa}).



Using the strong gravitational lensing power of rich galaxy clusters is a novel avenue to explore high-redshift galaxies (\citealt{Soucail:1990}). Galaxy candidates at $z \gtrsim 6$ that are magnified by foreground clusters were starting to be identified from \HST images more than a decade ago (e.g., \citealt{Ellis:2001fz,Hu:2002hb,Kneib:2004fx}); these observations provide an alternative way to probe the faint-end of the luminosity function with shorter exposure time than in blank fields. Recently, the Cluster Lensing And Supernova survey with Hubble (CLASH; \citealt{Postman:2012ca}) program and \HST-GO-11591 (PI: Kneib) program observed 34 galaxy clusters. The ongoing Hubble Frontier Fields (HFF; PI: Lotz\footnote{\url{http://www.stsci.edu/hst/campaigns/frontier-fields}}) program, upon its complete execution, will obtain deep \HST ACS/WFC3-IR images in six galaxy cluster fields (four of them are in the CLASH sample). \cite{Bradley:2014gk} recently reported 262 $6 \lesssim z \lesssim 8$ galaxy candidates across 18 clusters in the CLASH sample based on photometric redshift selection, and they demonstrated the power of using strong gravitational lensing to identify high-$z$ galaxies, especially at the bright end of the luminosity function. The even deeper HFF data, despite being $\gtrsim 0.7$ mag shallower than the HUDF data, can probe galaxies \emph{intrinsically} fainter than can be probed in the HUDF due to the power of gravitational lensing. Other coordinated campaigns are also underway to complement the deep \HST images in those targeted galaxy cluster fields (e.g., the Grism Lens-Amplified Survey from Space program; \citealt{Schmidt:2014kx,2015ApJ...812..114T}).

Here we use the deep \spitzer/IRAC images obtained from the \spitzer UltRa Faint SUrvey Program (SURFS UP; \citealt{Bradac:2014el}{; }{hereafter Paper I}) to probe the rest-frame optical emission from $z \gtrsim 6$ galaxy candidates. SURFS UP surveys 10 strong-lensing cluster fields (our sample partially overlaps with both CLASH and HFF) with \spitzer IRAC images in the $3.6\mu m$ and $4.5\mu m$ channels, with exposure times of $\gtrsim 28$ hours per channel per cluster. Paper I summarizes the science motivations and observational strategies of SURFS UP, and \cite{RyanJr:2014fu} presented the $z \sim 7$ galaxy candidates in the Bullet Cluster, one of which is detected in both IRAC channels. In this work, we explore the $6 \lesssim z \lesssim 10$ galaxy candidates with IRAC detections in 8 additional clusters in our sample\footnote{The \HST WFC3/IR imaging data for the tenth cluster, MACS2214, will be obtained in late 2015 (\HST-GO 13666; PI: Brada\v{c}).} and present their physical properties inferred from their broadband fluxes. We also make all the IRAC imaging data available for the community on our webpage\footnote{\url{http://www.physics.ucdavis.edu/∼marusa/SurfsUp.html}}. 

The structure of this paper is as follows: Section \ref{sec:data_photom} describes our \HST and \spitzer imaging data and photometry; Section \ref{sec:SampleSelection} describes our $6 \lesssim z \lesssim 10$ galaxy sample selection procedure; Section \ref{sec:spectroscopy} describes the identification of Ly$\alpha$ emission from spectroscopy for two galaxy candidates with IRAC detection at $z=6.76$ (RXJ1347-1216) and $z=6.32$ (MACS0454-1251); Section \ref{sec:sedfitting} presents our spectral energy distribution (SED) modeling procedure and results, and Section \ref{sec:irac_colors} explores the idea of using IRAC $[3.6]-[4.5]$ color to identify galaxies with strong nebular emission lines. Finally, Section \ref{sec:summary} summarizes our findings. Throughout the paper, we assume a $\Lambda$CDM concordance cosmology with $\Omega_m=0.3$, $\Omega_\Lambda=0.7$, and the Hubble constant $H_0 = 70\,\text{km}\,\text{s}^{-1}\,\text{Mpc}^{-1}$. Coordinates are given for the epoch J2000.0, and all magnitudes are in the AB system.

\section{Imaging Data and Photometry}\label{sec:data_photom}
 

\subsection{\HST Data and Photometry}\label{subsec:hst_data}

\begin{deluxetable*}{lccrrccc}
\tabletypesize{\scriptsize}
\tablecaption{SURFS UP Galaxy Cluster Sample\label{tab:cluster_sample}}
\tablecolumns{8}
\tablehead{\colhead{} &
           \colhead{Cluster Name} & 
           \colhead{Short Name\tablenotemark{a}} & 
           \colhead{R.A.} & 
           \colhead{Decl.} & 
           \colhead{$z_{\text{cluster}}$\tablenotemark{b}} & 
           \colhead{$N_{\text{LBG}}$\tablenotemark{c}} & 
           \colhead{$N_{\text{LBG,\,IRAC}}$\tablenotemark{d}} \\
           \colhead{} & 
           \colhead{} & 
           \colhead{} & 
           \colhead{(deg.)} & 
           \colhead{(deg.)} & 
           \colhead{} & 
           \colhead{} & 
           \colhead{} }

\startdata
1 & MACSJ0454.1$-$0300 & MACS0454 & 73.545417 & $ -3.018611 $ & 0.54 & 10 & 2 \\
2 & MACSJ0717.5$+$3745\tablenotemark{e,f} & MACS0717 & $109.390833$ & $ 37.755556 $ & 0.55 & 10 & 0 \\
3 & MACSJ0744.8$+$3927\tablenotemark{e} & MACS0744 & $116.215833$ & $ 39.459167 $ & 0.70 & 4 & 1 \\
4 & MACSJ1149.5$+$2223\tablenotemark{e,f} & MACS1149 & $177.392917$ & $ 22.395000 $ & 0.54 & 11 & 3 \\
5 & RXJ1347$-$1145\tablenotemark{e} & RXJ1347 & $206.883333$ & $ -11.761667 $ & 0.59 & 9 & 3 \\
6 & MACSJ1423.8+2404\tablenotemark{e} & MACS1423 & $215.951250$ & $ 24.079722 $ & 0.54 & 9 & 6 \\
7 & MACSJ2129.4$-$0741\tablenotemark{e} & MACS2129 & $ 322.359208 $ & $ -7.690611 $ & 0.59 & 0 & 0\\
8 & RCS2$-$2327.4$-$0204 & RCS2327 & $ 351.867500 $ & $ -2.073611 $ & $0.70$ & 6 & 1 \\
9 & 1E0657$-$56 & Bullet Cluster & $104.614167$ & $-55.946389$ & $0.30$ & 10 & 1 \\
10 & MACS2214.9$-$1359\tablenotemark{g} & MACS2214 & $333.739208$ & $-14.003000$ & $0.50$ & N/A & N/A \\
\hline
Total &  &  &  &  &  & 69 & 17
\enddata
\tablenotetext{a}{We will refer to each cluster by its short name.}
\tablenotetext{b}{Cluster redshift}
\tablenotetext{c}{Number of $\mathbf{6 \lesssim z \lesssim 10}$ LBG candidates selected by their \HST colors.}
\tablenotetext{d}{Number of $\mathbf{6 \lesssim z \lesssim 10}$ LBG candidates with $\geq 3\sigma$ detections in at least one IRAC channel.}
\tablenotetext{e}{A CLASH cluster}
\tablenotetext{f}{A Hubble Frontier Fields cluster}
\tablenotetext{g}{The \HST WFC3/IR data for MACS2214 will be collected in late 2015.}

\end{deluxetable*}

We list the eight galaxy clusters analyzed in this work in Table \ref{tab:cluster_sample}. Among the eight clusters, six (MACS0717, MACS0744, MACS1149, RXJ1347, MACS1423, and MACS2129) are in the Cluster Lensing And Supernova Survey (CLASH; \citealt{Postman:2012ca}) sample; therefore, each of them has \HST imaging data in at least twelve ACS/WFC and WFC3/IR filters\footnote{For the CLASH clusters, the ACS filters include F435W, F475W, F606W, F625W, F775W, F814W, and F850LP; the WFC3/IR filters include F105W, F110W, F125W, F140W, and F160W.} from the ACS and WFC3/IR cameras. The typical $5\sigma$ depths for the CLASH clusters reported by \cite{Postman:2012ca} are between $27.0$ and $27.5$ mag (within $0\farcs 4$ diameter apertures) for each filter, and the large number of filters provides unique constraints for high-$z$ galaxy searches among the \HST deep fields. 

In addition to the CLASH data, full-depth images for MACS0717 and MACS1149 from the Hubble Frontier Fields (HFF) program have also been released in June 2015. In these two clusters, \HST spent a total $\sim$140 orbits that are roughly split between ACS and WFC3/IR filters, and these images achieve $\approx 28.7\mbox{--}29\mbox{ mag}$\footnote{The magnitude limits are from the Frontier Fields website.} in the optical (ACS) and NIR (WFC3). We use the deepest HFF images for MACS1149 and MACS0717 for photometry in the filters where such images are available\footnote{The HFF filter sets include F435W, F606W, F814W, F105W, F125W, F140W, and F160W.}. The six CLASH clusters in our sample are also observed by the Grism Lens-Amplified Survey from Space (GLASS; PI: Treu; \citealt{Schmidt:2014kx,2015ApJ...812..114T}) program (MACS0717, MACS1149, MACS1423, RXJ1347, MACS0744, and MACS2129) and have deep grism spectra available. 

For the remaining two clusters being analyzed in this work, we have data for RCS2327 as part of the SURFS UP HST observations ({\HST}-GO13177 PI Brada\v{c}; \citealt{Hoag:2015}) and previous archival data ({\HST}-GO10846 PI Gladders; see also \citealt{Sharon:2015}). For MACS0454, we use the archival observations from {\HST}-GO11591 (PI: Kneib), GO-9836 (PI: Ellis), and GO-9722 (PI: Ebeling). We list the limiting magnitudes of point sources within a $0\farcs 4$ aperture for MACS0454 and RCS2327 in Table \ref{tab:maglim}.

\begin{deluxetable*}{cccccccccc}
\tabletypesize{\scriptsize}
\tablecaption{\HST $5\sigma$ Limiting magnitudes (point source, $0\farcs 4$ aperture) for RCS2327 and MACS0454\label{tab:maglim}}
\tablecolumns{10}
\tablehead{\colhead{Cluster} &
           \colhead{F435W} &
           \colhead{F555W} &
           \colhead{F775W} &
           \colhead{F814W} &
           \colhead{F850LP} &
           \colhead{F098M} & 
           \colhead{F110W} &
           \colhead{F125W} & 
           \colhead{F160W}}
\startdata
MACS0454 & \nodata & 27.7 & 26.9 & 27.9 & 26.5 & \nodata & 28.1 & \nodata & 27.4 \\
RCS2327  & 27.6 & \nodata & \nodata & 27.6 & \nodata & 27.3 & \nodata & 27.6 & 27.5 
\enddata

\end{deluxetable*}

We will use the template-fitting software \tphot (\citealt{Merlin:2015bw})  ---  the successor to TFIT (\citealt{Laidler:2007iy}) --- to measure the colors between \HST and \spitzer IRAC images (see Section \ref{subsec:irac_data}). To prepare the \HST images for \tphot, we use the public $0\farcs 03$/pixel CLASH images and block-sum the images to make $0\farcs 06$/pixel images. We also edit the astrometric image header values (\verb|CRVALs| and \verb|CRPIXs|) to conform to \tphot's astrometric requirements\footnote{\tphot requires that the pixel boundaries of the high- and low-resolution images be perfectly aligned, meaning that both images should have the same CRVALs and both have half-integer CRPIXs.} and make sure that \HST and \spitzer images are aligned to well within 0\farcs1 (Paper I).

We also match the point-spread functions (PSFs) among all \HST filters to get consistent colors. To do so, we identify isolated point sources in each cluster field, and we use the \texttt{psfmatch} task in \texttt{IRAF} to match all \HST images to have the same PSF as the reddest band, F160W. In practice, because of the small field of view of each cluster and the crowded environment, we can only select $\sim$ 5 isolated point sources in each cluster for PSF-matching. However, we measure the curves of growth of each point source after PSF-matching and find that in most filters, the curves match to within $\sim$20\% of that of F160W.

After pre-processing, we extract photometry on \HST images using SExtractor (\citealt{Bertin:1996ww}; version 2.8.6). We use the combined IR images as the detection image, and the SExtractor detection/deblending settings similar to (but slightly more conservative than) the values adopted by CLASH for their high-$z$ galaxy search (\citealt{Postman:2012ca}). Because our focus in this work is on identifying IRAC-detected high-$z$ sources, the slightly more conservative settings do not reject many potential IRAC-detected candidates but eliminates most spurious detections. We also follow the procedure outlined in \cite{Trenti:2011ft} to rescale the flux errors reported by SExtractor. At the end of the process, we use the resulting photometric catalogs and segmentation maps for both IRAC photometry (Section \ref{subsec:irac_data}) and color selection (Section \ref{sec:SampleSelection}). 




\subsection{IRAC Data and Photometry}\label{subsec:irac_data}

The IRAC imaging data set for SURFS UP was presented in Paper I; the total exposure time for each IRAC channel is about $30$ hours (see \citealt{RyanJr:2014fu} and Paper I for details.) The coadded IRAC mosaics are deeper within the main cluster fields covered by WFC3/IR, and the typical $3\sigma$ limiting magnitude within $3"$-radius apertures is $26.6$ mag in the $3.6$ $\mu$m channel (hereafter ch1) and $26.2$ mag in the $4.5$ $\mu$m channel (hereafter ch2) where source blending is not severe. 


We use \tphot to measure consistent colors between \HST and \spitzer IRAC images with a template-fitting approach (see also \citealt{Laidler:2007iy} for the template-fitting concept employed by \tphot.)  The template-fitting approach has been demonstrated to work well for blank-field surveys such as CANDELS (\citealt{Guo:2013ig}), but it does require images with zero mean background. Most galaxy cluster fields have considerable spatial variations in local sky background, so subtracting a constant background does not generally work. Therefore, instead of fitting all sources in the field at the same time --- which is the strategy for blank-field surveys --- we subtract the local background and perform the fit for each high-$z$ candidate separately to get the cleanest residual possible. 

As described in \cite{Merlin:2015bw}, \tphot is designed to measure the fluxes in the low-resolution image (in our case the IRAC images) for all the sources detected in the high-resolution image (in our case the F160W images). \tphot does so by constructing a template for each source; it convolves the cutout of each source in the F160W image by a PSF-transformation kernel that matches the F160W resolution to the IRAC resolution. Once the templates are available (and with their fluxes normalized to 1), \tphot solves the set of linear equations and finds the combination of coefficients for each template that most closely reproduce the pixel values in IRAC images; each coefficient is therefore the flux of the source in IRAC. \tphot also calculates the full covariance matrix and uses the diagonal terms of the covariance matrix to calculate flux errors. For each source, \tphot reports a ``covariance index'', defined as the ratio between the maximum covariance of the source with its neighbors (max($\sigma_{ij}$)) over its flux variance ($\sigma_{ii}$), which serves as an indicator of how strongly correlated the source's flux is with its closest (or brightest) neighbor. Generally, a high covariance index ($\gtrsim 1$) is associated with more severe blending and large flux errors, at least from simulations (\citealt{Laidler:2007iy,Merlin:2015bw}). Therefore, sources with high covariance indices should be treated with caution.

\begin{figure}[t]
\vspace*{-0.4in}
\includegraphics[width=\columnwidth]{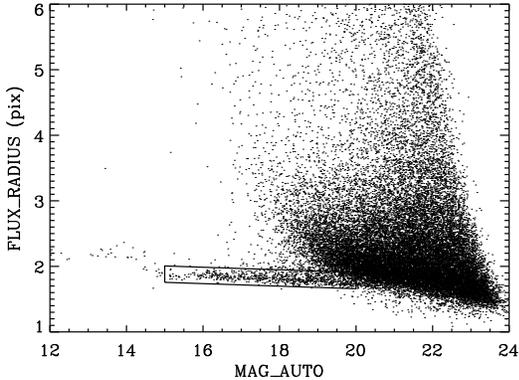}
\vspace*{-2in}
\caption{Brightness and half-light radii for all sources in IRAC ch1 of the SURFS UP clusters. The half-light radii (\texttt{FLUX\_RADIUS}) are in $0\farcs 6$ IRAC pixels. The box illustrates the crudely defined stellar loci where all point sources are expected to fall. When selecting putative point sources for each field, we first make sure the objects are near this locus for a given cluster, then place additional constraints on proximity to neighbors, axis ratio, and re-centering/alignment accuracy for the final sample per cluster. There are typically $\gtrsim 40$ stars per cluster for PSF generation. \label{fig:psf_locus}}
\end{figure}

Obviously the PSF-transformation kernel that matches the F160W PSF to the IRAC PSF is a crucial element in this process. We generate IRAC PSFs by stacking point sources observed in the exposures from both the primary cluster field and the flanking field. We identify point sources using Sextractor with \texttt{DEBLEND\_MINCONT}=$10^{-7}$, \texttt{MINAREA}=$9$, \texttt{DETECT\_THRESH}=\texttt{ANALYSIS\_THRESH}=$2$, and a Gaussian convolution kernel with $\sigma=3$ pixels (defined over a 5 by 5-pixel grid).  We require that all point sources have an axis ratio of $b/a > 0.9$, lie on the stellar locus within the box shown in Figure \ref{fig:psf_locus}, and are sufficiently separated from neighboring objects to have reliable centroids (\texttt{FLAGS}$\leq 3$).  We recompute the PSF centroids by fitting a Gaussian profile to the inner profile ($r<4$ pixels) using the Sextractor barycenters as initial guesses, and align the point sources using sinc interpolation. To mask neighboring objects, we grow the segmentation maps from Sextractor by 2 pixels.  At each phase we subtract the local sky (assuming there are no local gradients) and normalize the flux of the point source to unity.  We sigma-clip average the masked, registered, normalized point sources and do one further background correction only to ensure the convolutions with \tphot are flux conserving. As discussed in Paper I, our stacked PSFs are consistent with the IRAC handbook.  Each of our clusters contains at least 40 point sources per bandpass in our PSF-making process.

In practice, \tphot still breaks down in very crowded regions (e.g., near the cluster center or near bright cluster galaxies); in this case, we are limited mostly by our knowledge of the IRAC PSFs and our ability to subtract sky background underneath the sources. We also measure the ``reduced $\chi^2$'' for each source in IRAC within a $2\farcs4$ by $2\farcs4$ box by calculating the average difference per pixel between the model pixel values and the observed pixel values: $\chi^2_\nu = \sum_i^{N_{\mathrm{pix}}} (f_{i,\mathrm{model}} - f_{i,\mathrm{obs}})^2 / (f_{i,\mathrm{obs}}^2 \times N_{\mathrm{pix}})$, where $f_{i,\mathrm{model}}$ and $f_{i,\mathrm{obs}}$ are the model (best-fit) and observed flux in pixel $i$, respectively. Later in this work, we only report the IRAC fluxes of the high-$z$ galaxies with reliable IRAC flux measurements, i.e., $\chi^2_{\nu} \leq 3$.

For the sources with nominal $S/N$ above 3 but with poor \tphot residuals, we do not trust the \tphot-measured fluxes and estimate the local $3\sigma$ flux limits via artificial source simulations. We insert artificial point sources into the F160W image (but \textit{not} into the IRAC image) within 5" of the high-$z$ candidates and run \tphot to measure the local sky level. We repeat this process at least $100$ times near each high-$z$ candidate and use the resulting IRAC flux distribution to determine the $3\sigma$ flux limits. 
In our analysis in Section 5, we use the $3\sigma$ flux limit for only one IRAC filter for one source (ch1 for MACS1423-1384); for all the other sources, their IRAC flux measurements in both IRAC channels pass the $\chi^2_{\nu}$ test.

We also run a separate set of simulations that independently estimate the magnitude errors in case \tphot underestimates the magnitude errors in crowded regions even when they pass the $\chi^2_{\nu}$ test. In this set of simulations, we insert fake point sources around each high-$z$ target (within 5") in IRAC images with the same magnitude as the \tphot-measured value, and measure the flux of the fake sources again with \tphot. We then calculate the median difference between the input and output magnitudes of the fake sources as an independent magnitude error estimate. We find that for most sources, the \tphot-reported magnitude error is within 0.1 mag from the simulated magnitude error, but sometimes the simulated magnitude error is much larger than the \tphot-reported value. In these cases, \tphot might underestimate the true magnitude errors, so we adopt the simulated magnitude errors in our SED modeling. 
We note that by adding fake sources in IRAC images, we increase the flux error due to crowding, so the simulated magnitude errors could be higher than the true magnitude errors.

\section{Sample Selection}\label{sec:SampleSelection}

We select galaxy candidates at $z \gtrsim 6$ based on their rest-frame UV colors using the Lyman-break selection method (\citealt{Steidel:1993cd,Giavalisco:2002cv}). For the CLASH clusters, we use the published color criteria presented below for selecting $z \sim 6$, $7$, $8$, and $9$ Lyman-break galaxies (LBGs); for RCS2327 and MACS0454, we design our own color cuts. All galaxy colors are calculated using their isophotal magnitudes (\verb|MAG_ISO|) from SExtractor. After the initial color selection, we inspect the galaxy candidates to remove image artifacts and objects with problematic photometry. We then measure each LBG candidate's fluxes in IRAC, and we only present the candidates with $\mbox{S/N}\geq 3$ in at least one channel. Becuase of the differences in the available filters in each cluster, we explain our color-selection process in more detail below and list the sample size in Table \ref{tab:cluster_sample}. The full sample of the color-selected $z\gtrsim 6$ galaxy candidates with IRAC detections is presented in Table \ref{tab:irac_det_sample}.

\begin{deluxetable*}{lrrcccccrc}[ht]
\tabletypesize{\scriptsize}
\tablecaption{IRAC-Detected $6 \lesssim z \lesssim 10$ Galaxy Candidates\label{tab:irac_det_sample}}
\tablecolumns{9}
\tablehead{\colhead{Object ID} & 
           \colhead{R.A.} & 
           \colhead{Decl.} & 
           \colhead{F160W\tablenotemark{1}} & 
           \colhead{[3.6]\tablenotemark{2}} & 
           \colhead{$R_{3.6}$\tablenotemark{3}} & 
           \colhead{[4.5]\tablenotemark{4}} &
           \colhead{$R_{4.5}$\tablenotemark{3}} &
           \colhead{[3.6]$-$[4.5]} &
           \colhead{Spectroscopy{\tablenotemark{5}}} \\
           \colhead{} & 
           \colhead{(deg.)} & 
           \colhead{(deg.)} & 
           \colhead{(mag)} & 
           \colhead{(mag)} & 
           \colhead{} &
           \colhead{(mag)} &
           \colhead{} &
           \colhead{(mag)} &
           \colhead{}}

\startdata

%

\cutinhead{F125W-dropouts $(z \sim 9)$}
MACS1149-JD\tablenotemark{a} & $177.389943$ & $22.412719$ & $25.6 \pm 0.1$ & $25.8 \pm 0.4$ & $0.27$ & $25.0 \pm 0.2$ & $0.29$ & $0.8\pm 0.4$ & M,G\\


\cutinhead{F105W-dropouts $(z \sim 8)$}
MACS1423-1384 & $ 215.942115 $ & $ 24.079401 $ & $25.7 \pm 0.2$ & $> 23.6$\tablenotemark{b} & $0.95$ & $24.1 \pm 0.4 $\tablenotemark{c} & $0.95$ & $>-0.5$ & G \\
RXJ1347-1080\tablenotemark{h} & $206.891236$ & $-11.752594$ & $26.3 \pm 0.2$ & $25.4 \pm 0.2$ & $0.12$ & $25.3 \pm 0.2$ & $0.11$ & $0.2 \pm 0.2$ & D,G\\

\cutinhead{F850LP-dropouts $(z \sim 7)$}
MACS0744-2088\tablenotemark{h} & $ 116.250405 $ & $ 39.453011 $ & $25.5 \pm 0.2$ & $25.2 \pm 0.4 $\tablenotemark{c} & $0.50$ & $25.1 \pm 0.2$ & $0.50$ & $0.1 \pm 0.3$ & G \\
MACS1423-587 & $215.940493$ & $24.090848$ & $25.3 \pm 0.1$ & $24.3 \pm 0.3$\tablenotemark{c} & $0.86$ & $26.2 \pm 0.5$ & $0.80$ & $-1.8 \pm 0.6$  & G \\
MACS1423-774 & $215.935607$ & $24.086475$ & $25.9 \pm 0.2$ & $25.1 \pm 0.2$ & $0.70$ & $25.5 \pm 0.3$ & $0.69$ & $-0.4 \pm 0.3$ & D,G\\
MACS1423-2248 & $215.932958$ & $24.070875$ & $25.6 \pm 0.1$ & $25.0 \pm 0.1$ & $0.42$ & $25.3 \pm 0.2$ & $0.45$ & $-0.2 \pm 0.2$ & D,G\\
MACS1423-1494 & $215.935871$ & $24.078414$ & $26.3 \pm 0.2$ & $26.1 \pm 0.4$ & $0.92$ & $25.2 \pm 0.2$ & $0.89$ & $0.9 \pm 0.4$ & D,G\\
MACS1423-2097\tablenotemark{d} & $215.945534$ & $24.072433$ & $25.8 \pm 0.2$ & $24.6 \pm 0.3$\tablenotemark{c} & $0.68$ & $24.6 \pm 0.1$ & $0.68$ & $0.0 \pm 0.2$ & D,G\\
RXJ1347-1216\tablenotemark{d,e,h} & $206.900848$ & $-11.754199$ & $26.1 \pm 0.2$ & $24.3 \pm 0.1$ & $0.16$ & $25.6 \pm 0.2$ & $0.11$ & $-1.3 \pm 0.2$ & D,G\\
RXJ1347-1800 & $206.881657$ & $-11.761483$ & $25.4 \pm 0.2$ & $24.3 \pm 0.3$ & $0.63$ & $25.6 \pm 0.7$ & $0.55$ & $-1.3 \pm 0.8$ & G \\
Bullet-3\tablenotemark{g} & $104.667375$ & $-55.968067$ & $25.0 \pm 0.2$ & $23.8 \pm 0.3$ & n/a & $23.8 \pm 0.3$ & n/a & $0.0 \pm 0.4$ & FORS2 \\

\cutinhead{F814W-dropouts $(z \sim 6$--$7)$}
RCS2327-1282 & $351.880595$ & $-2.076292$ & $24.8 \pm 0.1$ & $24.4 \pm 0.1$ & $0.08$ & $24.1 \pm 0.1$ & $0.06$ & $0.3 \pm 0.1$ & D,M \\
MACS0454-1251\tablenotemark{f} & $73.535653$ & $-3.004116$ & $24.1 \pm 0.1$ & $23.2 \pm 0.2$ & $0.60$ & $23.4 \pm 0.2$ & $0.60$ & $-0.2 \pm 0.2$ & D \\
MACS0454-1817 & $73.551806$ & $-3.001018$ & $26.4 \pm 0.2$ & $24.1 \pm 0.3$\tablenotemark{c} & $0.31$ & $24.5 \pm 0.2$\tablenotemark{c} & $0.31$ & $-0.4 \pm 0.1$ & D \\

\cutinhead{F775W-dropouts $(z \sim 6)$}
MACS1149-274\tablenotemark{h} & $ 177.412009 $ & $ 22.415783 $ & $24.8 \pm 0.04$ & $24.1 \pm 0.1$ & $0.44$ & $24.0 \pm 0.1$ & $0.36$ & $0.0 \pm 0.1$ & G \\
MACS1149-1204\tablenotemark{h} & $ 177.378959 $ & $ 22.402429 $ & $25.0 \pm 0.1$ & $24.3 \pm 0.1$ & $0.64$ & $24.4 \pm 0.1$ & $0.67$ & $-0.1 \pm 0.1$ & G \\

\enddata

\tablenotetext{1}{Lensed total magnitude (MAG\_AUTO) in F160W; the magnification factors ($\mu$) are listed in Table \ref{tab:sed_results}.}
\tablenotetext{2}{Isophotal lensed magnitude in IRAC channel 1 based on the isophotal aperture defined in F160W.}
\tablenotetext{3}{$R_{3.6}$ and $R_{4.5}$ are the covariance indices for the $[3.6]$ and $[4.5]$ measurements, respectively. The covariance index of a source $i$ is defined as the ratio between the \textit{maximum} covariance among the neighbors ($\sigma_{ij}$) over the flux variance of itself ($\sigma_{ii}$) in the covariance matrix.}
\tablenotetext{4}{Isophotal lensed magnitude in IRAC channel 2 based on the isophotal aperture defined in F160W.}
\tablenotetext{5}{Instruments we used for spectroscopy: D=DEIMOS, M=MOSFIRE, G=\HST grism from the GLASS program (\citealt{Schmidt:2014kx}; Treu et al. 2015, in preparation)}
\tablenotetext{a}{First reported by \cite{Zheng:2012hu}.}
\tablenotetext{b}{The IRAC residual in the $3.6\mu$m channel shows that \tphot breaks down due to severe blending, so we report the simulated $3\sigma$ magnitude limit.} 
\tablenotetext{c}{\tphot likely underestimates the magnitude errors for these sources due to crowding, so we use the simulated magnitude errors.}
\tablenotetext{d}{Also reported by \cite{Smit:2014cg}.}
\tablenotetext{e}{Has Ly$\alpha$ detection at $z=6.76$.}
\tablenotetext{f}{Has tentative Ly$\alpha$ detection at $z=6.32$.}
\tablenotetext{g}{Reported by \cite{RyanJr:2014fu}.}
\tablenotetext{h}{Also reported by \cite{Bradley:2014gk}.}
\end{deluxetable*}

\subsection{CLASH \& Hubble Frontier Fields Clusters: MACS0717, MACS1149, MACS0744, MACS1423, MACS2129, and RXJ1347}
For the six clusters that in the CLASH sample, we use the criteria below. To be selected as a $z \sim 6$ LBG candidate, a source has to satisfy \emph{all} of the following criteria from \cite{Gonzalez:2011dn}:

\begin{equation}\label{eq:f775wDropsG11}
   \left\{
   \begin{aligned}
   &\text{F775W} - \text{F850LP} > 1.3\ \\ 
   &\text{F850LP} - \text{F125W} < 0.8\ \\
   &S/N \geq 5 \quad \text{in F850LP and F125W}\\\ 
   &S/N < 2    \quad \text{in filters bluer than F606W}
   \end{aligned} 
   \right.
\end{equation}
where we calculate all signal-to-noise ratios ($S/N$) within the isophotal (\verb|ISO|) aperture\footnote{The $S/N$ limits in blue filters roughly correspond to magnitude limits of $\gtrsim 28$ mag, based on the typical limiting magnitudes presented by \cite{Postman:2012ca}.}. If the $S/N$ in F775W is below one, we use the $1\sigma$ flux limit in F775W to calculate the $\text{F775W} - \text{F850LP}$ color. To select LBG candidates at $z \sim 7$, we use the color criteria from \cite{Bouwens:2011fe}:

\begin{equation}\label{eq:f850lpDropsB11}
   \left\{
   \begin{aligned}
   &\text{F850LP} - \text{F105W} > 0.7\ \\
   &\text{F105W} - \text{F125W} < 0.45\ \\
   &\text{F850LP}-\text{F105W} > 1.4 \times (\text{F105W} - \text{F125W}) + 0.42\\
   &S/N \geq 5\quad \text{in F105W and F125W}\\
   &S/N < 5\quad \text{in F814W}\ \\
   &S/N < 2\quad \text{in filters bluer than F775W}
   \end{aligned}
   \right.
\end{equation}

To select the LBGs at $z \sim 8$, we use the criteria from \cite{Bouwens:2011fe}:

\begin{equation}\label{eq:f105wDropsB11}
   \left\{
   \begin{aligned}
   &\text{F105W}-\text{F125W} > 0.45\ \\ 
   &\text{F125W}-\text{F160W}<0.5\ \\
   &S/N \geq 5\ \text{in F125W and F160W}\ \\
   &S/N < 2\ \text{in filters bluer than F850LP.}
   \end{aligned}
   \right.
\end{equation}

Finally, to select the LBG candidates at $z \gtrsim 9$, we use the criteria from \cite{Zheng:2014jn}:

\begin{equation}\label{eq:f125wDropsZ14}
   \left\{
   \begin{aligned}
   &\text{F125W}-\text{F160W} > 0.8\ \\
   &S/N \geq 5\ \text{in F160W}\ \\
   &S/N < 2\ \text{in filters bluer than F850LP.}
   \end{aligned}
   \right.
\end{equation}

In total, we find a total of 43 $z \gtrsim 6$ LBG candidates from the six CLASH/HFF clusters, among them 13 have $\geq 3\sigma$ detections in at least one IRAC channel.


\subsection{MACS0454}

\begin{figure}[t]
\vspace*{-1.0in}
\includegraphics[width=\columnwidth]{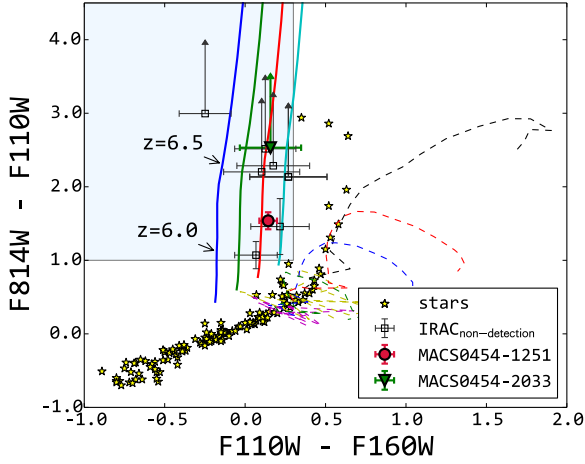}
\vspace*{-1.0in}
\caption{Color-color diagram of the F814W-dropout selection from MACS0454. The shaded region (light blue) shows where the expected F814W-dropout colors should be. We also plot (in \emph{solid} curves) the theoretical color tracks of a 100 Myr old stellar population with constant star formation taken from \cite{Bruzual:2003ck} with different amounts of dust attenuation. Sources within the shaded region that also pass the S/N cuts are shown in unfilled squares; two of them, MACS0454-1251 and MACS0454-1817 (shown in large filled symbols), are detected in IRAC. The color tracks of $z \leq 3$ galaxies, calculated from the local galaxy templates of \cite{Coleman:1980ic}, are shown in \emph{dashed} curves. We also show the expected colors of stars from \cite{Pickles:1998er}.}\label{fig:f814wDropsMACS0454}
\end{figure}

For the two galaxy clusters that are not in the CLASH sample (MACS0454 and RCS2327), we design our own selection criteria for $z \gtrsim 6$ galaxy candidates. For MACS0454, we have \HST imaging data in F555W, F775W, F814W, F850LP, F110W, and F160W, although the images in F775W and F850LP are shallower than the other filters and we use both filters only for $S/N$ rejection of low-$z$ interlopers. We use the following criteria to select $6 \lesssim z \lesssim 7.5$ galaxy candidates (F814W-dropouts):

\begin{equation}\label{eq:f814wDropsMACS0454}
   \left\{
   \begin{aligned}
   &\text{F814W}-\text{F110W} \geq 1.0\ \\ 
   &\text{F110W}-\text{F160W}\leq 0.3\ \\
   &S/N \geq 5\ \text{in F110W and F160W}\ \\
   &S/N < 2\ \text{in F555W.}
   \end{aligned}
   \right.
\end{equation}
We show the \HST color-color diagram of the F814W-dropout selection in Figure \ref{fig:f814wDropsMACS0454}. A total of ten sources satisfy the color and S/N cuts listed above; among them, MACS0454-1251 and MAC0454-1817 are detected in at least one IRAC channel. 

Because we have only one filter redward of F110W, we refrain from searching for F110W-dropouts in MACS0454 as it would yield objects detected in only one \HST filter.

\begin{figure}[t]
\vspace*{-1.0in}
\includegraphics[width=\columnwidth]{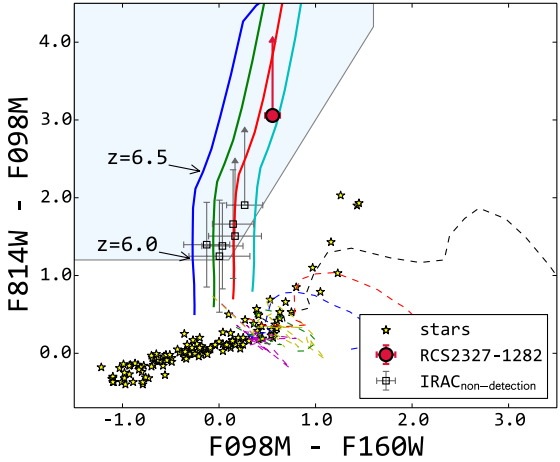}
\vspace*{-1.0in}
\caption{Color-color diagram of the F814W-dropout selection from RCS2327. The plot style and model assumptions are the same as Figure \ref{fig:f814wDropsMACS0454}. A total of six sources satisfy the F814W-dropout color criteria, and one of them (RCS2327-1282; shown as a red circle) has IRAC detections. \label{fig:f814wDropsRCS2327}}
\end{figure}

\subsection{RCS2327}\label{subsec:rcs2327}

For RCS2327, we have deep \HST images in F435W, F814W, F098M, F125W, and F160W, so we use the following criteria for $6 \lesssim z \lesssim 7.5$ galaxy candidates (F814W-dropouts):

\begin{equation} \label{eq:f814wDropsRCS2327}
   \left\{
   \begin{aligned}
   &\text{F814W}-\text{F098M} \geq 2.2\ \\ 
   &\text{F098M}-\text{F160W} \leq 1.6\ \\
   &\text{F814W}-\text{F098M} \geq 2.0 \times (\text{F098M}-\text{F160W}) + 1.0\ \\
   &S/N \geq 5\ \text{in\ F098M\ and\ F160W}\ \\
   &S/N < 3\ \text{in\ F435W.}
   \end{aligned}
   \right.
\end{equation}

When the $S/N$ in F814W is less than unity, we use its $1\sigma$ flux limit to calculate colors. We demonstrate the F814W-dropout selection from RCS2327 in Figure \ref{fig:f814wDropsRCS2327}, and six sources pass the above color and signal-to-noise cuts. One of the six sources, RCS2327-1218, is detected in both IRAC channels. In addition to the above criteria, we also use the criteria similar to the BoRG survey (\citealt{Trenti:2011ft}) to search for $z \gtrsim 7.5$ galaxy candidates:

\begin{equation}\label{eq:f098mDropsRCS2327}
   \left\{
   \begin{aligned}
   &\text{F098M}-\text{F125W} \geq 1.75\ \\
   &S/N \geq 5\ \text{in\ F125W\ and\ F160W}\ \\
   &S/N < 3\ \text{in\ F814W\ and\ F435W.}
   \end{aligned}
   \right.
\end{equation}
and when the $S/N$ in F098M is less than unity, we use its $1\sigma$ flux limit to calculate colors. The F098M-dropout search yields no galaxy candidate, so we find a total of one galaxy candidate with IRAC detections at $z \gtrsim 6$ from RCS2327 (see also \citealt{Hoag:2015} for more details on the dropout search in RCS2327).

\begin{figure*}[t]
\includegraphics[width=\textwidth]{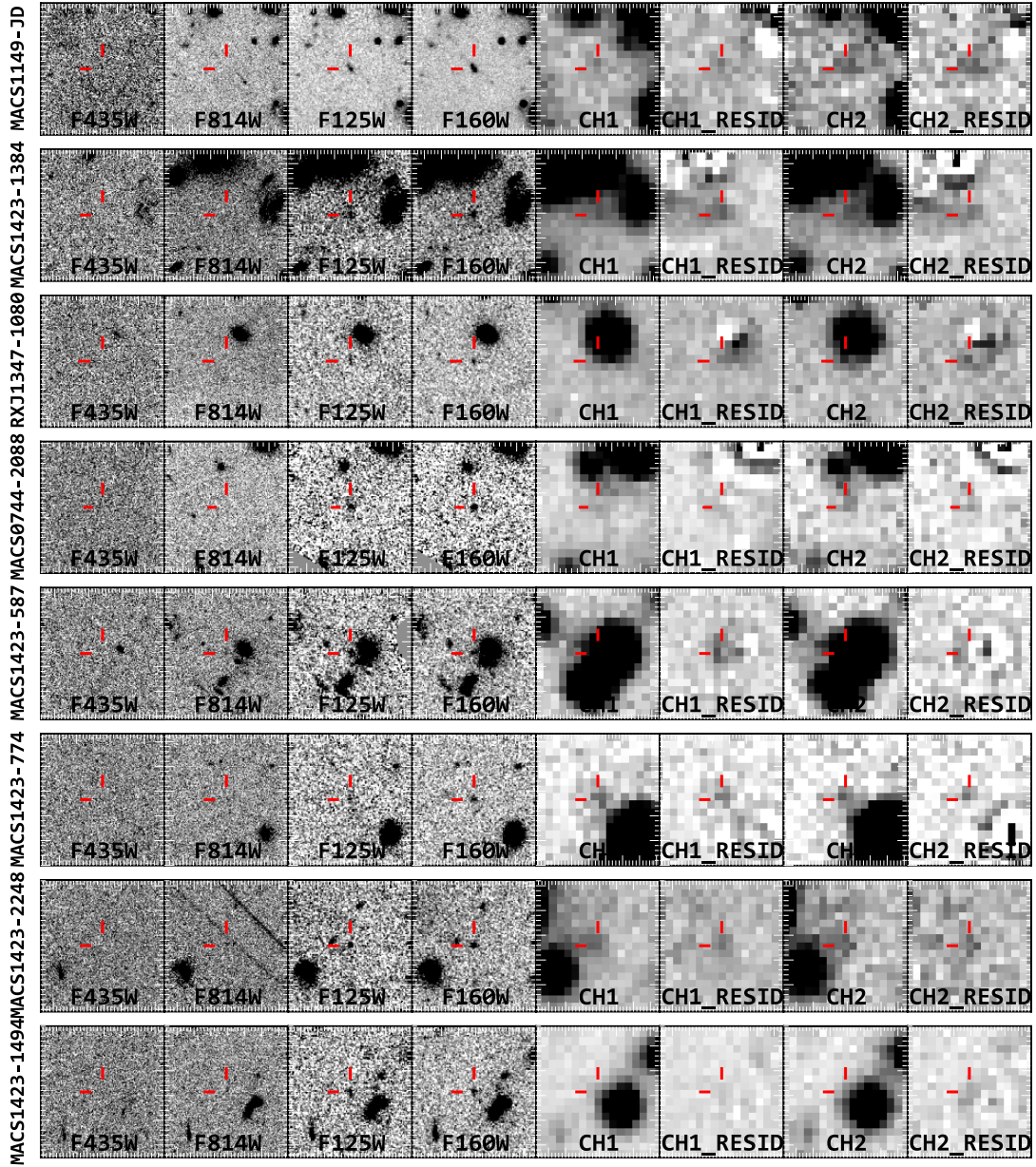}
\vspace*{-1in}
\caption{Cutouts of the first eight IRAC-detected, $z \gtrsim 6$ LBG candidates in SURFS UP (excluding the one candidate in the Bullet Cluster reported by \citealt{RyanJr:2014fu}). Each row shows the cutout in two \HST ACS filters (F435W and F814W in all clusters except for MACS0454, where we show F435W and F555W), two \HST WFC3/IR filters (F125W and F160W), and two IRAC channels. We also show the neighbor-subtracted cutouts around each LBG candidate in both IRAC channels (designated by \texttt{CH1\_RESID} and \texttt{CH2\_RESID}, respectively). The LBG candidate ID is to the left of each row. Each panel is centered on the LBG candidate (marked by the red lines), and each panel spans 10" by 10" on the sky.}\label{fig:montage}
\end{figure*}

To summarize, we find a total of 69 LBG candidates at $z \gtrsim 6$ from 9 clusters in SURFS UP; 17 of them have IRAC detections in at least one channel. Figure \ref{fig:montage} shows the cutouts of the 16 IRAC-detected LBG candidate in \HST and \spitzer images (one candiate was reported by \citealt{RyanJr:2014fu}). We also report the IRAC photometry for the entire sample in Table \ref{tab:irac_det_sample}. We use the simulated IRAC magnitude errors for MACS1423-1384, MACS1423-587, MACS0744-2088, MACS1423-2097, and MACS0454-1817, because we find that \tphot likely underestimates their IRAC magnitude errors from the simulations. We keep MACS1423-1384 in our sample because it has a nominal $5.9\sigma$ detection in ch2 from \tphot, although the simulated magnitude error suggests that the additional flux error due to crowding reduces it to a $2.2\sigma$ detection in ch2.


\begin{figure*}[t]
\includegraphics[width=\textwidth]{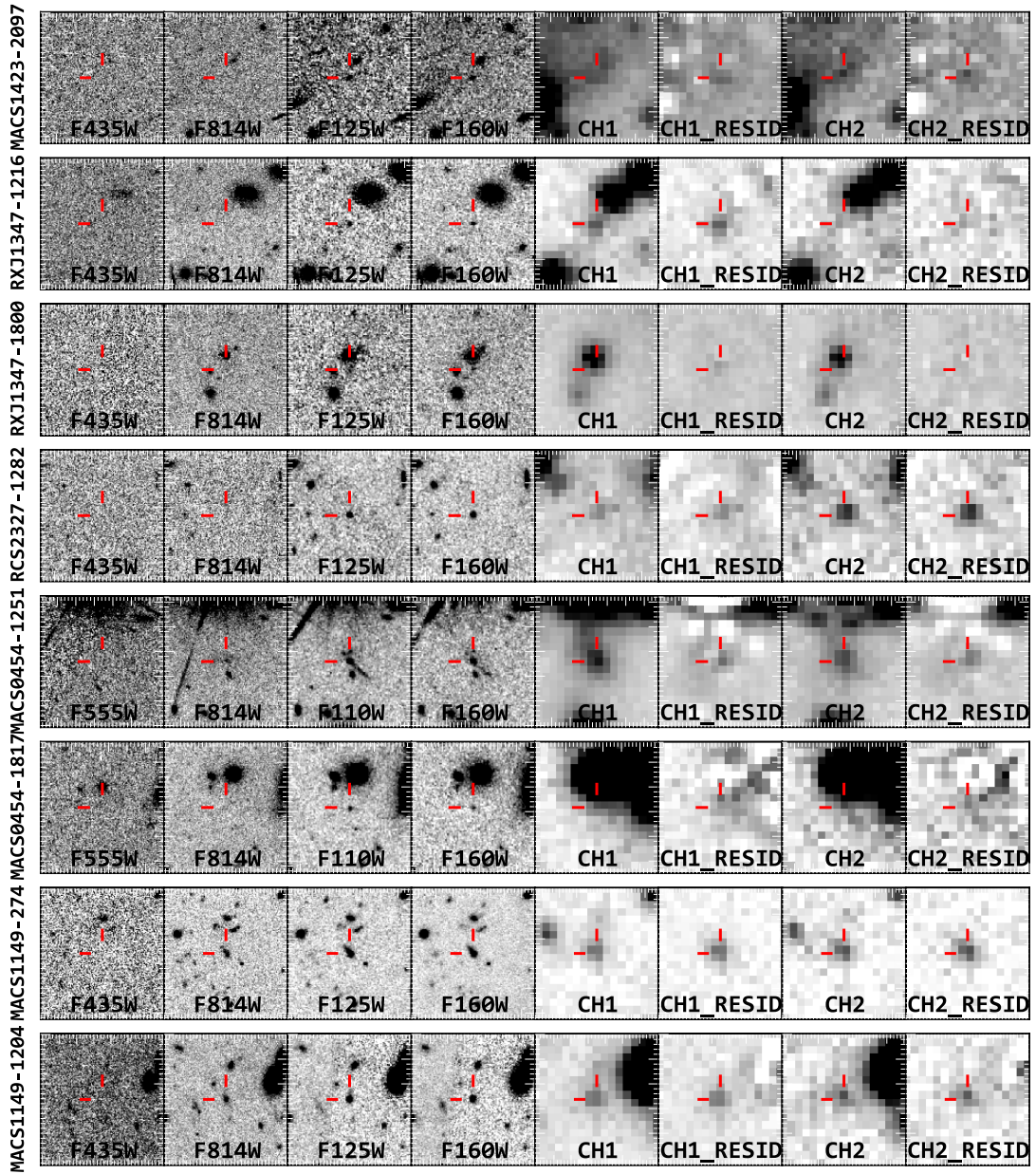}
\vspace*{-1in}
\caption{Same as Figure \ref{fig:montage}, but for the remaining seven LBG candidates.} \label{fig:montage2}
\end{figure*}
\clearpage  

\section{Spectroscopic Observations}\label{sec:spectroscopy}



We report the detection of two likely Ly$\alpha$ emitters among our sample with DEIMOS (\citealt{Faber:2003ev}) on the Keck II telescope.  The DEIMOS observation is part of a larger campaign to systematically target lensed high-$z$ galaxies behind strong-lensing galaxy clusters with DEIMOS and MOSFIRE (\citealt{McLean:2010ij,McLean:2012gv}) on Keck.
We observed six cluster fields between 2013 April and 2014 August and targeted 9 out of 17 high-$z$ galaxy candidates in Table \ref{tab:irac_det_sample} with DEIMOS. Which galaxy candidates were observed with DEIMOS and MOSFIRE are indicated in Table \ref{tab:irac_det_sample}. The DEIMOS data were reduced using the standard DEEP2 spec2d pipeline slightly modified to reduce the data observed also with 600 l/mm and 830 l/mm gratings (\citealt{Lemaux:2009fy,Newman:2013ha}).  We focus here on the two galaxies (RXJ1347-1216 and MACS0454-1251) that have line detections; we will present the full spectroscopic survey (with both DEIMOS and MOSFIRE) and the line flux limits for the non-detections in a future work. In addition to the Keck observations, 13 of the 17 galaxy candidates in Table \ref{tab:irac_det_sample} are also observed by the Grism Lens-Amplified Survey from Space (GLASS; HST GO-13459; PI: Treu) program; the spectroscopic constraints on Ly$\alpha$ emission from the \HST grism data will be presented by Schmidt et al., 2015 (in preparation).


Below we discuss the two galaxy candidates with robust line detections and the likelihood that they are Ly$\alpha$ emitters at $z=6.76$ and $z=6.32$.


\subsection{RXJ1347-1216}\label{subsec:rxj1347-1216}

\begin{figure}
\vspace{-1.3in}
\includegraphics[width=\columnwidth,trim=0 250 0 0,clip]{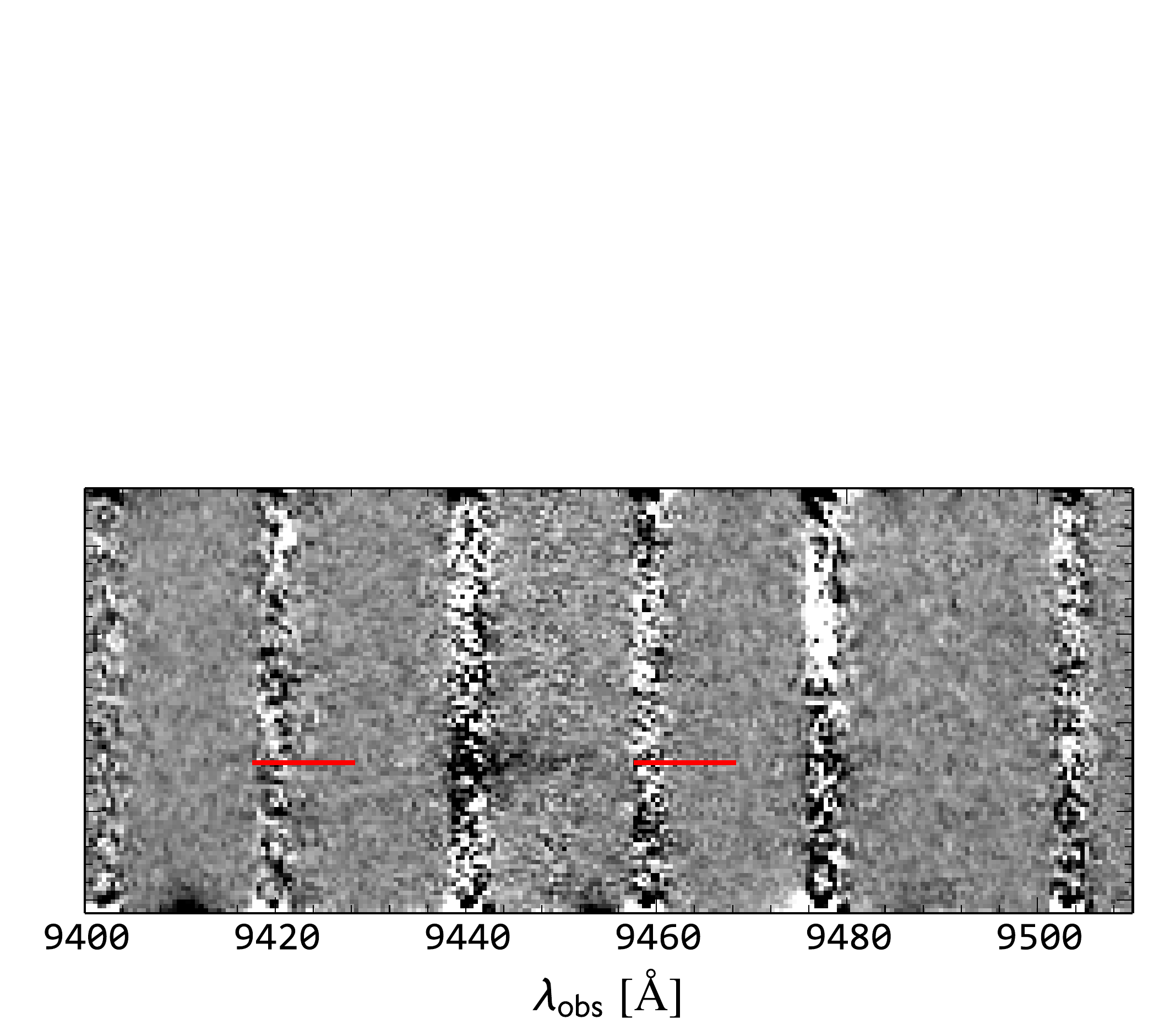}
\includegraphics[width=\columnwidth,trim=0 250 0 300,clip]{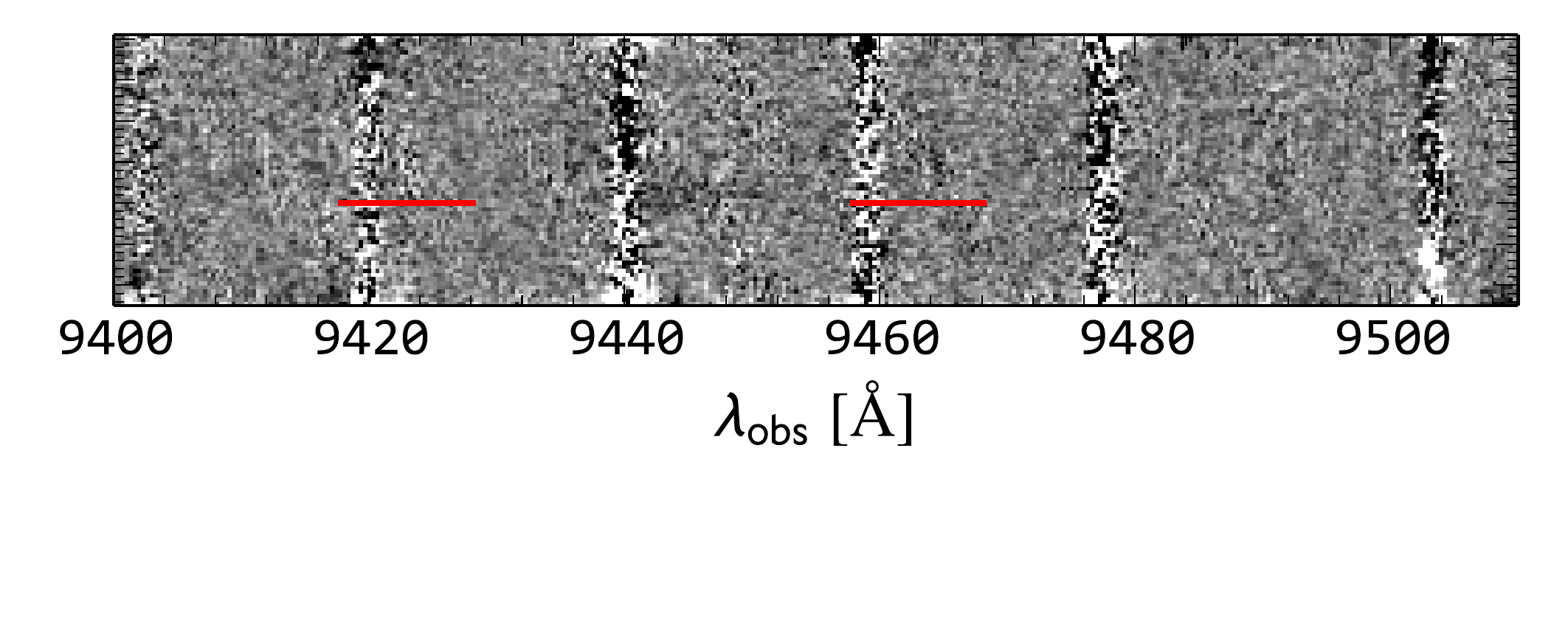}
\includegraphics[width=\columnwidth,trim=0 150 0 200]{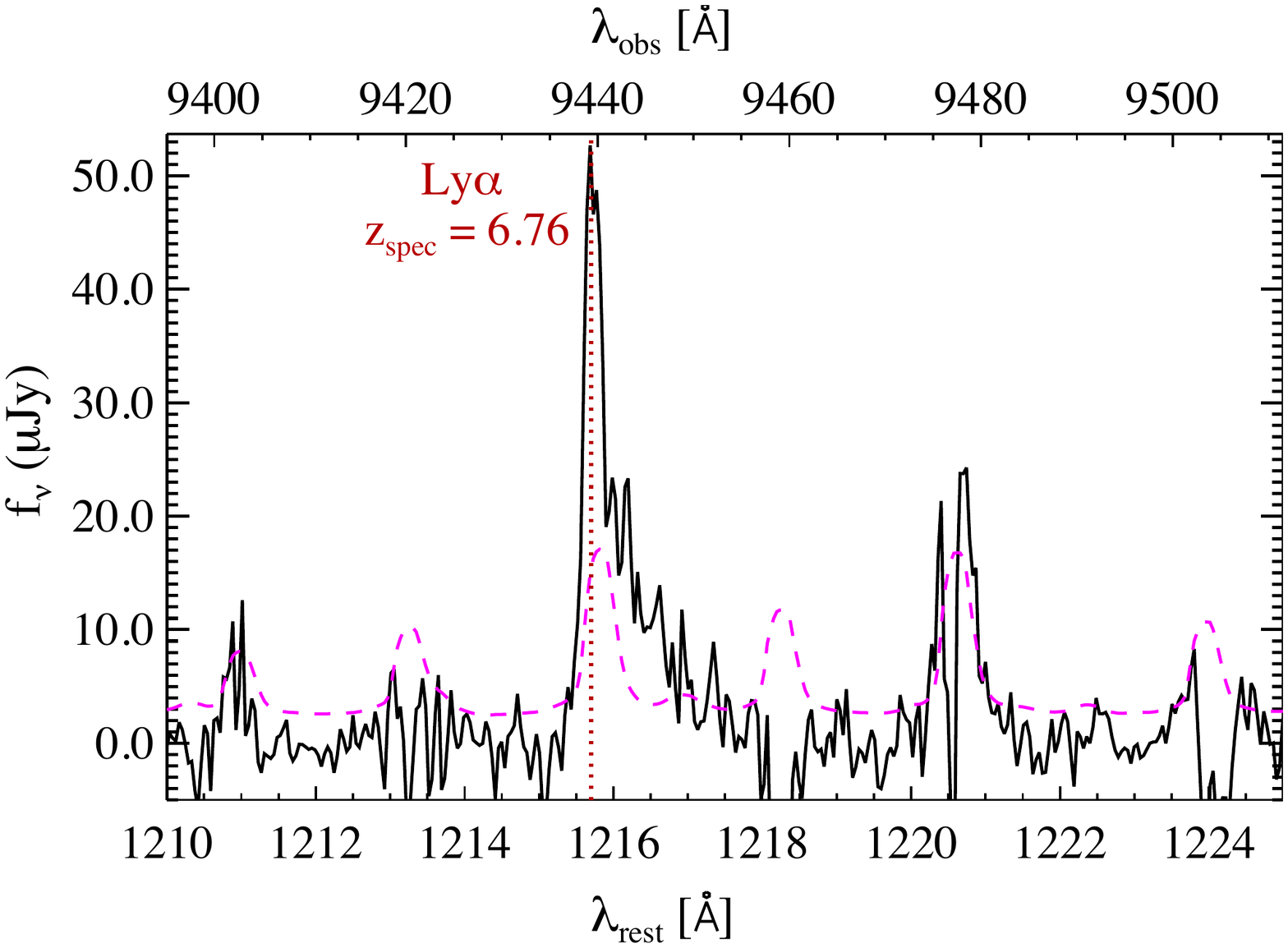}

\figcaption{Reduced two-dimensional (top and middle panels) and one-dimensional (bottom panel) spectrum of RXJ1347-1216. The top panel shows the data taken with the 830 l/mm grating on April 3, 2013, and the middle panel shows the data taken with the 1200 l/mm grating on May 26, 2014. The one-dimensional spectrum was extracted from the 830 l/mm spectrum. We also plot the RMS spectra in dashed lines and mark the emission line redshift if the line is to be Ly$\alpha$. The flux density values given on the ordinate are calculated in the rest-frame assuming the line to be Ly$\alpha$.\label{fig:spec_rxj1347_1216}}
\end{figure}

\begin{figure}
\vspace{-1.5in}
\includegraphics[width=\columnwidth,trim=0 250 0 0,clip]{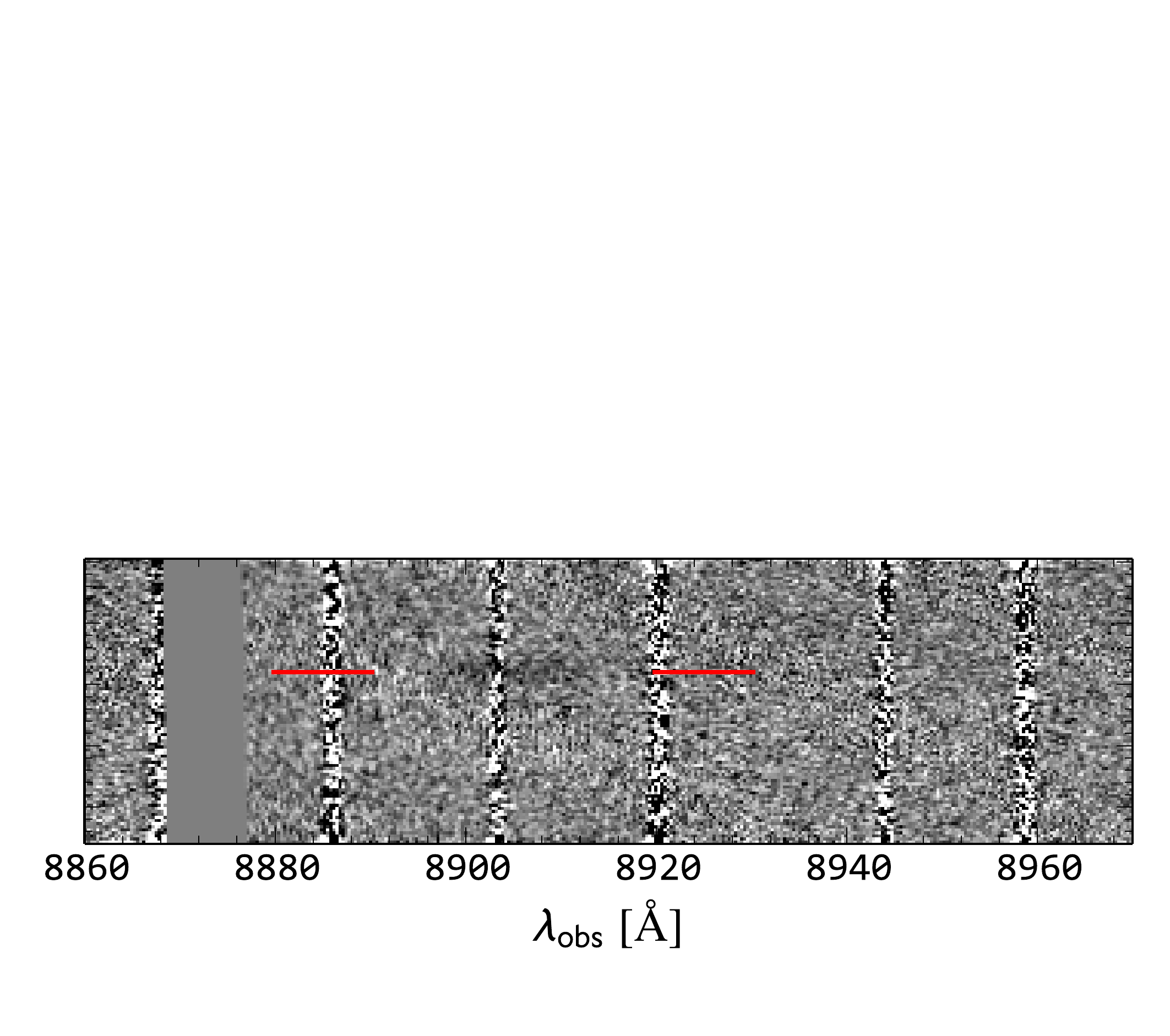}
\includegraphics[width=\columnwidth,trim=0 280 0 300,clip]{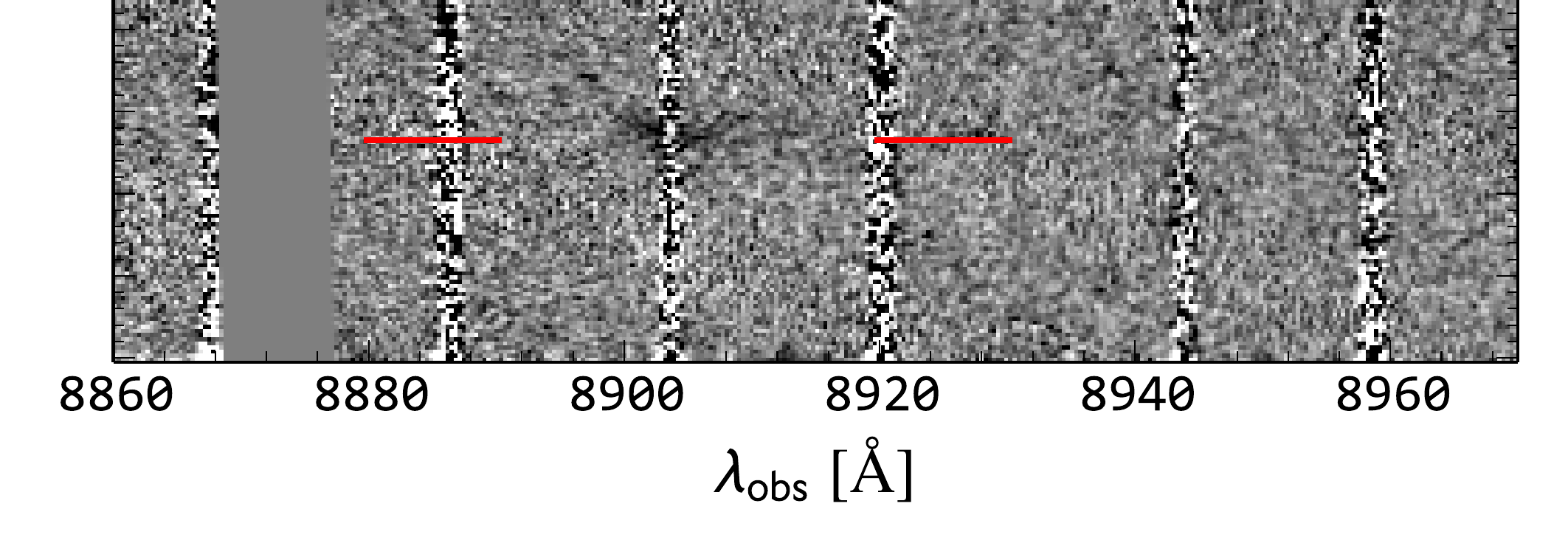}
\includegraphics[width=\columnwidth,trim=0 150 0 150,clip]{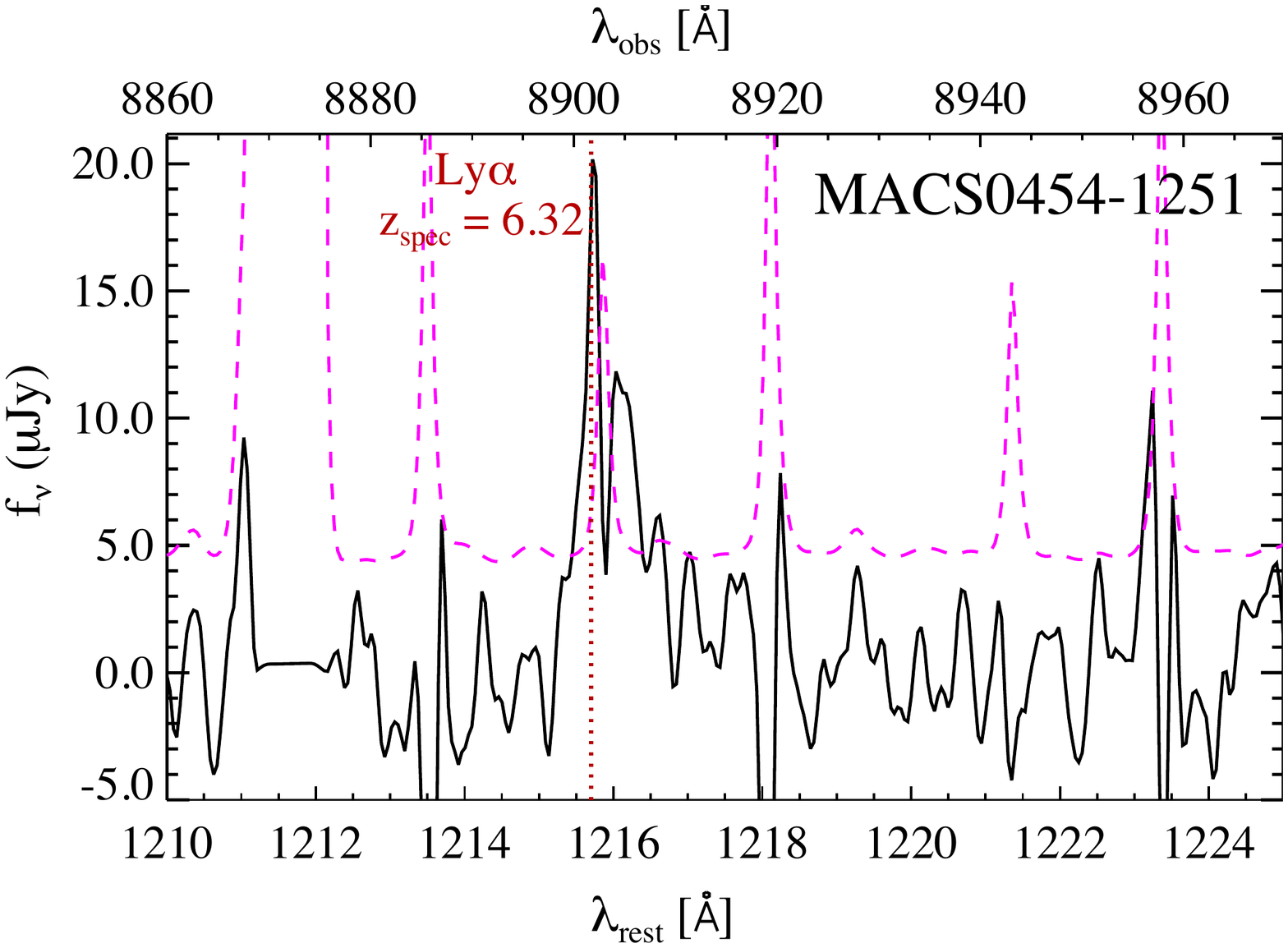}
\figcaption{Reduced two-dimensional (top and middle panels) and one-dimensional (bottom panel) spectrum of MACS0454-1251. The top panel shows the data taken on Nov 27, 2014, and the middle panel shows the data taken on Nov 28, 2014; we observed with the 1200 l/mm grating on both nights. The one-dimensional spectrum (bottom panel) was extracted from the data in the first night (top panel). We also plot the RMS spectra in dashed lines and mark the emission lines redshift if the line is to be Ly$\alpha$. The flux density values given on the ordinate are calculated in the rest-frame assuming the line to be Ly$\alpha$.\label{fig:macs0454_1251}}

\end{figure}

We selected this object as a $z \sim 7$ LBG candidate for spectroscopic follow-up on April 3 2013 and May 26 2014. This source was also selected by both \cite{Smit:2014cg} and \cite{Bradley:2014gk} as a $z_{\text{phot}} \sim 6.7$ LBG with high [OIII]$+$H$\beta$ equivalent widths ($>1300$\AA\ rest frame). We used the 830 l/mm grating in the 2013 run and the 1200 l/mm grating in the 2014 run, and the total integration times are roughly 6000 and 7200 seconds, respectively. We had good (but not photometric) conditions with $\sim 1\arcsec$ seeing in the 2013 run, but the conditions were highly variable in the 2014 run. Therefore, we only present the line flux measurements from the 2013 run, though the line was detected significantly in both runs.

In Figure \ref{fig:spec_rxj1347_1216} we show the two-dimensional spectra of RXJ1347-1216 from both observation runs (in the top and middle panels) and the combined one-dimensional spectrum (in the bottom panel). We detect an extended emission feature with FWHM$_{\text{obs}}$ $\sim$ 16.5\,\AA\ in the 2013 run, and although the blue side of the feature is severely contaminated by sky line residual, its asymetric profile with a tail to the red side of the spectrum strongly suggests that it is Ly$\alpha$. Using the centroid of the sky line residual at 9439\AA\ as the peak of of line profile, we determine its Ly$\alpha$ redshift as $z_{\text{Ly}\alpha}=6.76 \pm 0.003$ (the uncertainty corresponds to the width of the sky line residual). The Ly$\alpha$ redshift is in excellent agreement with its photometric redshift $z_{\text{phot}}=6.8\pm 0.1$, lending additional support to the identification of the Ly$\alpha$ feature. 

The emission feature is also independently detected at $4\sigma$ in the GLASS grism data at $\sim 9440$\AA. With the grism spectra in both G102 and G141, one can test the possibility of this feature being an [OII] doublet ($\lambda 3727$, $\lambda 3729$) at $z \sim 1.5$ by looking for the [OIII] $\lambda 5007$ line at $\sim 1.3\mu$m. For a typical star forming galaxy, the line ratio [OIII]/[OII] should be at least $\sim 0.3$ (\citealt{Jones:2015}), and the [OIII]/[OII] ratio is even higher for low-metallicity galaxies (e.g., \citealt{Maiolino:2008gs}). Assuming the detected line in DEIMOS is [OII] at $z=1.53$, \HST G141 grism data imply a $2\sigma$ upper limit on [OIII]/[OII] $\lesssim 0.3$, which is highly unlikely for a star forming galaxy (Schmidt et al., 2015). Therefore, we conclude that the \HST grism data also strongly support the $z=6.76$ Ly$\alpha$ interpretation of this emission line.

We perform line flux measurements from the DEIMOS data obtained during the 2013 run (with the 830 l/mm grating) following the procedure outlined in Section 2.4 of \cite{Lemaux:2009fy}. In short, we place two filters of width 20\AA\ on both sides of the emission line that are free of spectral features and sky line residuals to measure the background, and a central filter encompassing the emission line to measure the integrated flux. The background, which is fit to a linear function, is then subtracted from the integrated line flux. We perform spectrophotometric calibration of the DEIMOS data using two other compact sources (with half-light radii $\sim 0\farcs3$ as measured from the F775W image) on the same mask with continuum detection in the following manner: for each object, the combination of slit loss and loss due to clouds was determined by calculating a spectral magnitude, done by correcting each DEIMOS spectrum for spectral response and atmospheric extinction and convolving the F775W filter curve with the resulting spectra, and comparing this value with the magnitude measured in the \HST image. The ratio of the flux densities for each of the two sources is calculated and averaged to estimate the total spectral loss for this mask, which is then applied to RXJ1347-1216 assuming a similar half-light radius for this object. The reason for this assumption is, though the half-light radius of RXJ1347-1216 is smaller ($0\farcs2$), which would, in principle, mean less slit loss, the size of the Ly$\alpha$ nebula is known to far exceed the size of the UV continuum region (e.g., \citealt{Wisotzki:2015}). For the two sources on the mask with which we performed the flux calibration, the total measured throughput of the slit was $\sim$40\%, lower than the $\sim$60\% expected for sources of this size (if they are symmetric), suggesting at least some departure from a photometric night.


Using the above procedure, we measure the line flux from the 830 l/mm data to be $7.8 \pm 0.7 \times 10^{-18}\ \text{erg}\ \text{cm}^{-2}\ \text{s}^{-1}$, which translates into a Ly$\alpha$ luminosity of $4.1 \pm 0.4 \times 10^{42}\ \text{erg}\ \text{s}^{-1}$. We do not detect continuum in the spectrum, so we estimate the rest-frame Ly$\alpha$ equivalent width using the object's broadband magnitude in F105W (on the red side of Ly$\alpha$) to be $26 \pm 4$ \AA. The equivalent width uncertainty include the Poisson noise in the central filter encompassing the sky line residual, the uncertainty in the continuum, and the uncertainty in DEIMOS absolute flux calibration.

We note that our measured Ly$\alpha$ line flux from DEIMOS data is roughly a factor of 3 lower than that measured from the \HST grism data, which is $2.6 \pm 0.5 \times 10^{-17}\ \text{erg}\ \text{cm}^{-2}\ \text{s}^{-1}$ (Schmidt et al., 2015, ApJ, in press). The difference in the line flux measurements could be due to the following factors: (1) our DEIMOS data are shallower than the \HST grism data ($\sim 2$ hours of total integration time in the 2013 mask, versus $\sim 5$ orbits of \HST integration time in G102 grism), with the difference in depths leading to differences in the spatial and extent of the emission detected above the noise in each observation, which impacts the total integrated line flux; (2) we might still underestimate the Ly$\alpha$ slit loss because the true extent of the Ly$\alpha$-emitting region is much larger than the continuum-emitting region, while the \HST slitless grism recovers more of the Ly$\alpha$ flux; (3) there is a sky line coincident with the peak wavelength of the Ly$\alpha$ emission in the DEIMOS data which appears slightly over-subtracted that serves to slightly lower the measured line flux; and (4) there could also be issues with contamination subtraction in the \HST grism data, although it is unclear which direction it biases line flux measurement. We do expect the ground-based line flux measurement to be a lower limit to the space-based measurement, which is consistent with the measured values from DEIMOS and from \HST grism data.

We also measure the inverse line asymmetry $1/a_\lambda$ as defined in \cite{Lemaux:2009fy}, and the line's inverse asymmetry value ($1/a_\lambda=0.22$) is well within the range ($1/a_\lambda < 0.75$) typical for convincing Ly$\alpha$ emission.

Using Ly$\alpha$ line flux, one can also infer a star formation rate following \cite{Lemaux:2009fy}. The inferred star formation rate is strictly a lower limit, because our conversion assumes no attenuation of Ly$\alpha$ photons by dust or neutral hydrogen. The inferred star formation rate is $1.6 \pm 0.1\ M_{\odot}\ \mathrm{yr}^{-1}$, consistent with being a lower limit to the value we derive from SED fitting in Section \ref{sec:sedfitting}, $17.0 \pm 0.5\,\sfrunit$. The Ly$\alpha$-inferred star formation rate roughly yields a Ly$\alpha$ escape fraction of $\sim 10$\%.

\subsection{MACS0454-1251}\label{subsec:macs0454-1251}

We observed MACS0454 with DEIMOS on November 28 and 29, 2014 using the 1200 l/mm grating on both nights. The total exposure time for this mask is 7200s. We also reduce the DEIMOS data and extract one-dimensional spectrum following the procedure in \cite{Lemaux:2009fy}, in the same way as RXJ1347-1216.

We show the reduced two-dimensional spectra of MACS0454-1251 in Figure \ref{fig:macs0454_1251}, and an extended emission feature is clearly detected on both nights. However, a bright sky line residual cuts through the middle of the emission feature in the spectral direction and makes the line interpretation ambiguous. Given the width of the emission line, it could be either Ly$\alpha$ at $z = 6.32$ or [OII] at $z = 1.39$, but the sky line residual makes it difficult to either confirm or rule out Ly$\alpha$ or [OII] based on line shape alone. However, as we will show in Section \ref{subsec:model_results}, this line is more likely to be Ly$\alpha$ than [OII] because its \HST fluxes (and upper limits) are much better fit by a galaxy template at $z=6.32$ than at $z=1.39$ (Section \ref{subsec:model_results}), and its photometric redshift probability density function $P(z)$ has very a low probability at $z=1.39$ (see the $P(z)$ curve in Figure \ref{fig:sedfits_all2}). We measure the line fluxes from both nights to be $1.2 \pm 0.2 \times 10^{-17}\ \text{erg}\ \text{cm}^{-2}\ \text{s}^{-1}$ (first night) and $8.0 \pm 1.5 \times 10^{-18}\ \text{erg}\ \text{cm}^{-2}\ \text{s}^{-1}$ (second night), and these translate to Ly$\alpha$ luminosities of $5.5 \pm 0.9 \times 10^{42}\ \text{erg}\ \text{s}^{-1}$ (first night) and $3.6 \pm 0.3 \times 10^{42}\ \text{erg}\ \text{s}^{-1}$ (second night). From the line fluxes, we estimate its rest-frame equivalent widths (assuming Ly$\alpha$) using the continuum on the red side of Ly$\alpha$ (estimated from its F110W flux density) from both nights to be $8.2 \pm 1.4$ \AA\ and $5.4 \pm 1.0$ \AA. We also infer star formation rate lower limits to be $2.3 \pm 0.4\ M_{\odot}\ \mathrm{yr}^{-1}$ (first night) and $1.5 \pm 0.3\ M_{\odot}\ \mathrm{yr}^{-1}$ (second night) based on Ly$\alpha$ fluxes, also fully consistent with being lower limits to the value derived from SED fitting in Section \ref{sec:sedfitting}, $17.0^{+18.0}_{-4.1}\,M_{\odot}\,\mathrm{yr}^{-1}$. The Ly$\alpha$-inferred star formation rate also roughly corresponds to an escape fraction of $\sim 10$\% for Ly$\alpha$ photons.

The inverse asymmetry value measured for the one-dimensional emission feature is $\sim$0.5 for both masks, consistent with the values typical for Ly$\alpha$. However, the line asymmetry estimate is less reliable than that of RXJ1347-1216 because we have to mask the over-subtracted skyline near the central wavelength of the emission feature. Based on the object's photometric information and line asymmetry measurements, we identify this source as a Ly$\alpha$ emitter at $z = 6.32$, although we are less confident with this Ly$\alpha$ interpretation than we are with RXJ1347-1216. MACS0454 is not in the GLASS sample, so we are unable to cross check our measurements with \HST grism data.

\section{Redshift and Spectral Energy Distribution Modeling}\label{sec:sedfitting}

In this section, we present our stellar population modeling of the IRAC-detected, $z \gtrsim 6$ galaxy candidates using broadband photometry. We present the modeling procedure in Section \ref{subsec:model_procedure}, the modeling results in Section \ref{subsec:model_results}, and we discuss the sources of bias and uncertainty in Section \ref{subsec:sedfit_error}.


\subsection{Modeling Procedure}\label{subsec:model_procedure}
For photometric redshift and stellar population modeling, we use the photometric redshift code EAZY (\citealt{Brammer:2008gn}) with stellar population templates from \citet[BC03]{Bruzual:2003ck} with \cite{Chabrier:2003ki} initial mass function (IMF) between $0.1$ and $100\ M_\odot$ and a metallicity of $0.2\ Z_\odot$. There is very little direct observational evidence for galaxy metallicity at $z \geq 3$, but limited results so far suggest that the majority of them have sub-solar metallicity (\citealt{Maiolino:2008gs}); we choose $0.2\ Z_\odot$ for easy comparison with other works. The galaxy templates are generated assuming an exponentially declining star formation history with $e$-folding time $\tau$ ranging from $0.1$ and $30$ Gyr, and ages of the stellar population range from $10$ Myr to $13$ Gyr. For each combination of age and $\tau$, we implement galaxy internal dust attenuation using the \cite{Calzetti:2000iy} prescription, with the reddening parameter $E(B-V)$ ranging from $0$ to $1$ mag to include potential low-$z$ dusty galaxy solutions. The $E(B-V)$ grid we use have a step size of $\Delta E(B-V)=0.02$ mag from $E(B-V)=0$ to $0.5$ mag and a step size of $\Delta E(B-V)=0.1$ mag from $E(B-V)=0.5$ to $1$ mag. We also use the stellar templates from the photometric code \emph{Le Phare} (\citealt{Ilbert:2006bw})\footnote{The fit using stellar templates is still done using EAZY, only the templates are from \emph{Le Phare}.} in the fitting to check if stellar templates provide significantly better fits.

\begin{figure}[t]
\vspace{-1in}
\includegraphics[width=\columnwidth]{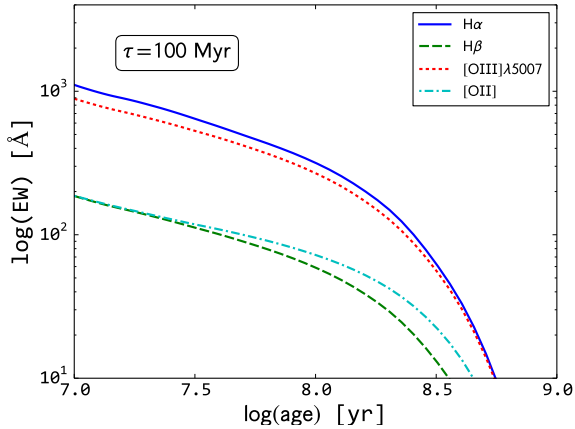}
\vspace{-1in}
\caption{Equivalent widths of H$\alpha$, H$\beta$, [OII] $\lambda\lambda 3726,\,3729$, and [OIII] $\lambda 5007$ that we add to the BC03 models as a function of stellar population age, assuming that all Lyman continuum photons are converted into nebular emission. The BC03 galaxy templates used in this figure have a metallicity of $0.2 Z_{\odot}$, no dust attenuation, and a star-formation rate $e$-folding time of $100$ Myr. Some galaxy candidates (e.g., RXJ1347-1216) require strong nebular emission lines to explain their observed IRAC colors. \label{fig:nebular_EW}}
\end{figure}

Recent studies have shown that for some galaxy candidates, strong nebular emission lines contribute significantly to broadband fluxes and therefore influence the inferred galaxy properties (e.g., \citealt{Schaerer:2010gc, Smit:2014cg}). Therefore, we use galaxy templates that include nebular emission lines in the modeling. In order to calculate the expected line fluxes for a given BC03 galaxy template, we calculate the integrated Lyman continuum flux (before dust attenuation) and use the relation from \cite{Leitherer:1995gx} to calculate the expected fluxes from hydrogen recombination lines (mainly H$\alpha$, H$\beta$, Pa$\beta$, and Br$\gamma$) while assuming the Lyman continuum escape fraction to be zero. Non-zero Lyman continuum escape fraction will reduce the strength of optical nebular emission lines (see \citealt{Inoue:2011cu} and \citealt{Salmon:2015iz}). We then use the tabulated line ratios between H$\beta$ and the metal lines from \cite{Anders:2003ci} to calculate the metal line fluxes for a metallicity of $0.2Z_{\odot}$.  For templates with dust attenuation, we include the dust attenuation effects \emph{after} adding nebular emission lines. The resultant equivalent widths as a function of galaxy age, for $\tau=100$ Myr and $E(B-V)=0$, are shown in Figure \ref{fig:nebular_EW}, in agreement with \cite{Leitherer:1995gx}.

In addition to nebular emission lines, we also include nebular \emph{continuum} emission that account for the bound-free emission of HI and HeI as well as the two-photon emission of hydrogen from the 2s level. We follow the prescription in \cite{Krueger:1995ty} (their equations 7 and 8) to calculate the nebular continuum flux as a function of Lyman continuum photon density, and we calculate the emission coefficients using the methods in \cite{Brown:1970gf} and \cite{Nussbaumer:1984ur}\footnote{The fitting formula in \cite{Nussbaumer:1984ur} is crucial to calculate the two-photon continuum emission between rest-frame 1216\AA\ and 2431\AA, where two-photon emission dominates the nebular continuum.}. Nebular continuum emission could be an important component for very young ($\sim 10$Myr) starbursts and can contribute up to $\sim 1/3$ of the total continuum just blueward of the rest-frame 4000\AA\ break. Nebular continuum emission also makes the rest-frame UV slope redder than expected from stars alone (\citealt{Schaerer:2010gc}).


We do not include Ly$\alpha$ in our galaxy templates. Strong Ly$\alpha$ emission could affect the LBG color selection by changing the rest-frame UV broadband colors and could affect the derived physical properties from SED fitting (\citealt{Schaerer:2011fr,DeBarros:2014fa}). But Ly$\alpha$ photons suffer from complicated radiative transfer processes and does not show tight correlations with the stellar population properties, so for simplicity we do not include Ly$\alpha$ emission in our modeling. 

Strong gravitational lensing boosts galaxy fluxes and increases the \emph{apparent} SFR and stellar mass. To calculate the \emph{unlensed} SFR and stellar mass, we use the magnification factor $\mu_{\text{best}}$ estimated from the cluster mass models for each galaxy candidate at its redshift. We generate our own models for MACS1149, MACS0717, MACS0454, RXJ1347, and RCS2327, following the procedures outlined in \cite{Bradac:2005ea,Bradac:2009bk}. In short, we constrain the gravitational potential on a mesh grid within a galaxy cluster field via $\chi^2$ minimization, and we adaptively use denser pixel grids near the core(s) of the cluster and around multiple images. We find the minimum $\chi^2$ values by iteratively solving a set of linearized equations that satisfy $\partial \chi^2 / \partial \psi_k = 0$, where $\psi_k$ is the gravitational potential in the $k$th dimension. We then produce the magnification ($\mu$) map from the best-fit gravitational potential map. For the rest of the clusters in our sample (MACS1423, MACS2129, and MACS0744), we use the public PIEMD-eNFW\footnote{PIEMD-eNFW models use pseudo-isothermal elliptical mass distributions for galaxies and elliptical NFW profiles for dark matter.} models by \cite{Zitrin:2012fo}.\footnote{These models are made public as high-end science products of the CLASH program; \url{http://archive.stsci.edu/prepds/clash/}.} The magnification factors (and their errors) are estimated at the galaxy candidate positions from the $z=9$ magnification maps except for MACS1423-587, MACS1423-774, MACS1423-2248, and RXJ1347-1800 (which have $z_{\text{best}}\sim 1$ so we estimate their $\mu_{\text{best}}$ from the $z = 1$ magnification map.)

\begin{deluxetable*}{llcccccccccr}[ht]  
\tabletypesize{\scriptsize}
\tablecaption{Photometric Redshift and Stellar Population Modeling Results\label{tab:sed_results}}
\tablecolumns{10}  
\tablehead{ \colhead{Object ID} &
            \colhead{$z_{\text{best}}$\tablenotemark{a}} &
            \colhead{$\mubest$\tablenotemark{b}} &
            \colhead{$M_{\text{stellar}}\times f_{\mu}$\tablenotemark{c}} &
            \colhead{SFR$\times f_{\mu}$\tablenotemark{c}} &
            \colhead{Age\tablenotemark{d}} &
            \colhead{sSFR\tablenotemark{e}} &
            \colhead{$E(B-V)_{\text{fit}}$\tablenotemark{h}} &
            \colhead{$\beta$\tablenotemark{i}} &
            \colhead{$M_{1600}-2.5\log(f_{\mu})$\tablenotemark{k}} \\
            \colhead{} &
            \colhead{} & 
            \colhead{} & 
            \colhead{($10^9\,M_{\odot}$)} &
            \colhead{($\sfrunit$)} &
            \colhead{(Myr)} & 
            \colhead{(Gyr$^{-1}$)} &
            \colhead{(mag)} &
            \colhead{} &
            \colhead{(mag)} }

\startdata
 
\cutinhead{F125W-dropouts ($z \sim 9$; $\mstar = -20.63 \pm 0.36$ at $z \sim 8$ from \citealt{Bouwens:2015gm})}

MACS1149-JD & $9.3^{+0.1}_{-0.1}$ & $5.5^{+0.3}_{-0.3}$ & $0.9^{+0.5}_{-0.4}$ & $1.9^{+0.6}_{-0.5}$ & $200^{+250}_{-60}$ & $2.1^{+3.1}_{-0.7}$ & $0.00$ & \nodata & $-19.9 \pm 0.1$ \\

\cutinhead{F105W-dropouts ($z \sim 8$; $\mstar = -20.63 \pm 0.36$ at $z \sim 8$ from \citealt{Bouwens:2015gm})}

RXJ1347-1080 & $7.3^{+0.3}_{-0.4}$ & $6.0^{+0.6}_{-0.6}$ & $0.6^{+0.2}_{-0.4}$ & $0.5^{+1.0}_{-0.5}$ & $290^{+350}_{-260}$ & $0.9^{+1.9}_{-0.9}$ & $0.00$ & $-2.5^{+1.2}_{-1.1}$ & $-18.9 \pm 0.2$ \\

MACS1423-1384 & $6.9^{+0.9}_{-0.1}$ & $4.0^{+1.2}_{-1.2}$ & $1.1^{+3.2}_{-0.6}$ & $115.6^{+16.1}_{-114.6}$ & $\leq 130$ & $105.0^{+0.1}_{-98.9}$ & $0.38$ & $-0.5^{+1.6}_{-0.6}$ & $-19.4 \pm 0.2$ \\

\cutinhead{F850LP-dropouts ($z \sim 7$; $\mstar = -20.87 \pm 0.26$ from \citealt{Bouwens:2015gm})}

MACS1423-1494 & $7.1^{+0.3}_{-0.5}$ & $1.7^{+0.1}_{-0.1}$ & $0.2^{+1.2}_{-0.1}$ & $22.1^{+4.3}_{-19.3}$ & $\leq 10$ & $105.1^{+0.1}_{-9.2}$ & $0.14$ & $-2.1^{+1.1}_{-1.4}$ & $-20.1 \pm 0.2$ \\

MACS0744-2088 & $7.0^{+0.2}_{-0.1}$ & $1.5^{+0.1}_{-0.1}$ & $0.7^{+1.2}_{-0.4}$ & $34.0^{+4.0}_{-28.1}$ & $20^{+180}_{-0}$ & $45.7^{+16.8}_{-42.3}$ & $0.16$ & $-0.8^{+0.7}_{-1.2}$ & $-20.7 \pm 0.1$ \\

RXJ1347-1216\tablenotemark{g} & 
   $6.76$ &
   $5.0^{+0.5}_{-0.5}$ &
   $0.2^{+0.1}_{-0.1}$ &
   $17.0^{+2.6}_{-2.7}$ &
   $\leq 10$ &
   $105.0^{+0.1}_{-0.1}$ &
   $0.20$ &
   $-2.5^{+0.7}_{-1.0}$ &
   $-19.3 \pm 0.1$
   \\

MACS1423-2097 & $6.8^{+0.1}_{-0.2}$ & $3.5^{+0.1}_{-0.1}$ & $2.9^{+0.3}_{-2.5}$ & $1.5^{+20.0}_{-0.1}$ & $720^{+90}_{-490}$ & $0.5^{+90.9}_{-0.1}$ & $0.00$ & $-0.6^{+0.6}_{-1.1}$ & $-19.7 \pm 0.1$ \\

Bullet-3\tablenotemark{l} & $6.8^{+0.1}_{-0.1}$ & $12^{+4.0}_{-4.0}$ & $2.0^{+0.6}_{-0.8}$ & $1.3^{+1.4}_{-0.6}$ & $630^{+160}_{-230}$ & ${0.7}^{+0.5}_{-0.6}$ & $0.00$ & \nodata & $-18.9 \pm 0.4$ \\

MACS1423-587 & $0.1^{+6.7}_{-0.1}$ & $1.5^{+0.1}_{-0.1}$ & $0.1^{+0.2}_{-0.1}$ & $\leq 0.1$ & $11500^{+1250}_{-11490}$ & $ \leq 0.1$ & $0.70$ & \nodata & \nodata \\

RXJ1347-1800 & $0.8^{+0.7}_{-0.5}$ & $4.1^{+0.1}_{-0.1}$ & $0.1^{+0.1}_{-0.1}$ & $\leq 1.2$ & $2600^{+7830}_{-2570}$ & $ \leq 36.0$ & $0.00$ & \nodata & \nodata \\

MACS1423-774 & $1.2^{+5.5}_{-0.3}$ & $1.2^{+0.1}_{-0.1}$ & $0.3^{+0.5}_{-0.2}$ & $\leq 20.6$ & $1430^{+180}_{-1420}$ & $ \leq 100.1$ & $0.06$ & \nodata & \nodata \\

MACS1423-2248 & $1.2^{+0.3}_{-1.0}$ & $1.2^{+0.1}_{-0.1}$ & $1.0^{+2.7}_{-1.0}$ & $\leq 5.2$ & $5000^{+3640}_{-4980}$ & $ \leq 52.9$ & $0.00$ & \nodata & \nodata \\

\cutinhead{F814W-dropouts ($z \sim 6-7$; $\mstar = -20.87 \pm 0.26$ from \citealt{Bouwens:2015gm})}

RCS2327-1282 & $7.7^{+0.1}_{-0.4}$ & $4.1^{+0.5}_{-0.4}$ & $1.9^{+0.4}_{-0.5}$ & $6.1^{+0.5}_{-0.2}$ & $400^{+100}_{-120}$ & $3.2^{+1.6}_{-0.4}$ & $0.00$ & $-3.0^{+0.2}_{-0.3}$ & $-20.9 \pm 0.1$ \\

MACS0454-1817 & $6.5^{+0.2}_{-0.1}$ & $2.6^{+0.3}_{-0.3}$ & $0.4^{+3.8}_{-0.1}$ & $38.6^{+28.9}_{-36.3}$ & $\leq 30$ & $105.0^{+0.1}_{-67.6}$ & $0.28$ & $-1.4^{+0.5}_{-0.8}$ & $-19.3 \pm 0.1$ \\

MACS0454-1251 & $6.1^{+0.1}_{-0.1}$ & $4.4^{+0.4}_{-0.4}$ & $2.1^{+2.8}_{-0.8}$ & $17.9^{+14.1}_{-5.8}$ & $90^{+40}_{-10}$ & $8.7^{+4.8}_{-2.5}$ & $0.12$ & $-1.6^{+0.2}_{-0.2}$ & $-20.8 \pm 0.1$ \\

MACS0454-1251\tablenotemark{j}  &  $6.32$ & 
   $4.4^{+0.4}_{-0.4}$ &
   $0.5^{+2.9}_{-0.2}$ &
   $19.0^{+7.9}_{-5.4}$ & 
   $30^{+0}_{-0}$ & 
   $39.6^{+65.5}_{-35.3}$ &
   $0.08$ &
   $-1.6^{+0.2}_{-0.2}$ &
   $-20.9 \pm 0.1$ 
   \\

\cutinhead{F775W-dropouts ($z \sim 6$; $\mstar = -20.94 \pm 0.20$ from \citealt{Bouwens:2015gm})}

MACS1149-274 & $5.8^{+0.1}_{-0.1}$ & $1.6^{+0.1}_{-0.1}$ & $2.4^{+0.5}_{-0.5}$ & $17.7^{+5.2}_{-0.1}$ & $100^{+10}_{-20}$ & $7.3^{+3.2}_{-1.2}$ & $0.08$ & $-1.6^{+0.1}_{-0.1}$ & $-21.2 \pm 0.1$ \\

MACS1149-1204 & $5.7^{+0.1}_{-0.1}$ & $1.8^{+0.1}_{-0.1}$ & $1.3^{+0.3}_{-0.8}$ & $12.9^{+4.7}_{-2.0}$ & $80^{+30}_{-50}$ & $10.2^{+29.2}_{-2.8}$ & $0.08$ & $-1.5^{+0.4}_{-0.5}$ & $-20.7 \pm 0.2$ \\

\enddata
\tablenotetext{a}{$\zbest$ is the photometric redshift using BC03 galaxy templates except for RXJ1347-1216 and MACS0454-1251, for which we identify Ly$\alpha$ emission at $z_{\mathrm{spec}}=6.76$ and $z_{\mathrm{spec}}=6.32$, respectively.}
\tablenotetext{b}{Lensing magnification factor estimated from the galaxy cluster mass models mentioned in Section \ref{subsec:model_procedure}. For MACS1423-587, MACS1423-774, MACS1423-2248, and RXJ1347-1800, we estimate their $\mu_{\text{best}}$ from the magnification map at $z = 1$.}
\tablenotetext{c}{The \emph{intrinsic} stellar mass and SFR assuming $\mu = \mubest$. To use a different magnification factor $\mu$, simply use $f_{\mu} \equiv \mu / \mubest$, where $\mubest$ is the best magnification factor we adopt for each object. When the best-fit SFR is zero, we report the 68\% upper limit.}
\tablenotetext{d}{Time since the onset of star formation. For the sources with best-fit age equal to 10 Myr, the youngest template allowed in our models, we report the 68\% upper limit of the age from Monte Carlo simulations.}
\tablenotetext{e}{Specific star formation rate $\equiv$ SFR / stellar mass. When the best-fit sSFR is zero, we report the 68\% upper limit.}
\tablenotetext{g}{All fits are performed at $z=6.76$, the Ly$\alpha$ redshift. The confidence intervals reflect the maximal range returned from the simulations because the distributions for this object are highly skewed.}
\tablenotetext{h}{The best-fit color excess $E(B-V)$ of the stellar emission from our SED modeling. The dust attenuation at rest-frame $1600$\AA\ can be calculated using the dust attenuation curve from \cite{Calzetti:2000iy} as $A_{1600} = 9.97 \times E(B-V)$.}
\tablenotetext{i}{Measured rest-frame UV slope $\beta$ from \HST/WFC3 broadband fluxes, assuming the candidates are at $z=z_{\text{best}}$. We do not measure $\beta$ for MACS1423-587, MACS1423-774, MACS1423-2248, and RXJ1347-1800 because they have $z_{\text{best}} \sim 1$. We also do not measure $\beta$ for MACS1149-JD because it has only 1 filter (F160W) that samples the rest-frame UV continuum.}
\tablenotetext{j}{All fits are performed at $z=6.32$, its Ly$\alpha$ redshift.}
\tablenotetext{k}{Rest-frame 1600 \AA\ absolute magnitude assuming $z=\zbest$ and $\mu=\mubest$.}
\tablenotetext{l}{First reported by \cite{RyanJr:2014fu}; included here for completeness.}
\end{deluxetable*}

\subsection{Modeling Results}\label{subsec:model_results}

\begin{figure*}[hp]
\centering
\vspace{-0.5in}
\includegraphics[width=\textwidth]{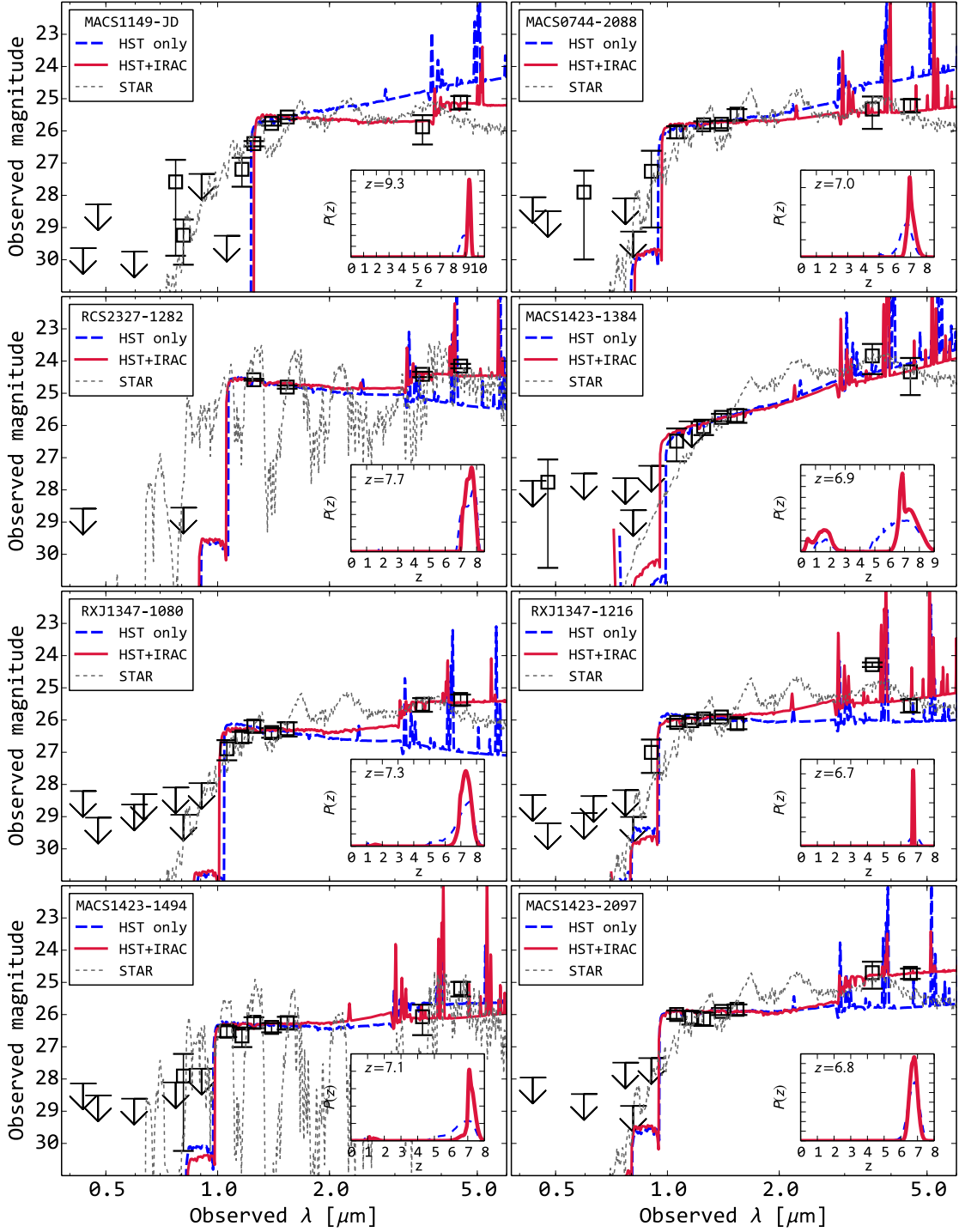}
\vspace{-0.5in}
\caption{Best-fit SEDs for each galaxy candidate with IRAC detections. The best-fit SEDs using only \HST photometry are shown in dashed blue lines, while the SEDs using combined \HST and IRAC photometry are shown in solid red lines. The best-fit stellar templates (fixed at $z=0$) are shown in thin dotted lines. The photometric redshift probability density functions $P(z)$ are shown as insets. The photometric redshifts decreases from top to bottom first, then from the left column to the right column. \label{fig:sedfits_all}}
\end{figure*}

\begin{figure*}[hp]
\centering
\vspace{-0.5in}
\includegraphics[width=\textwidth]{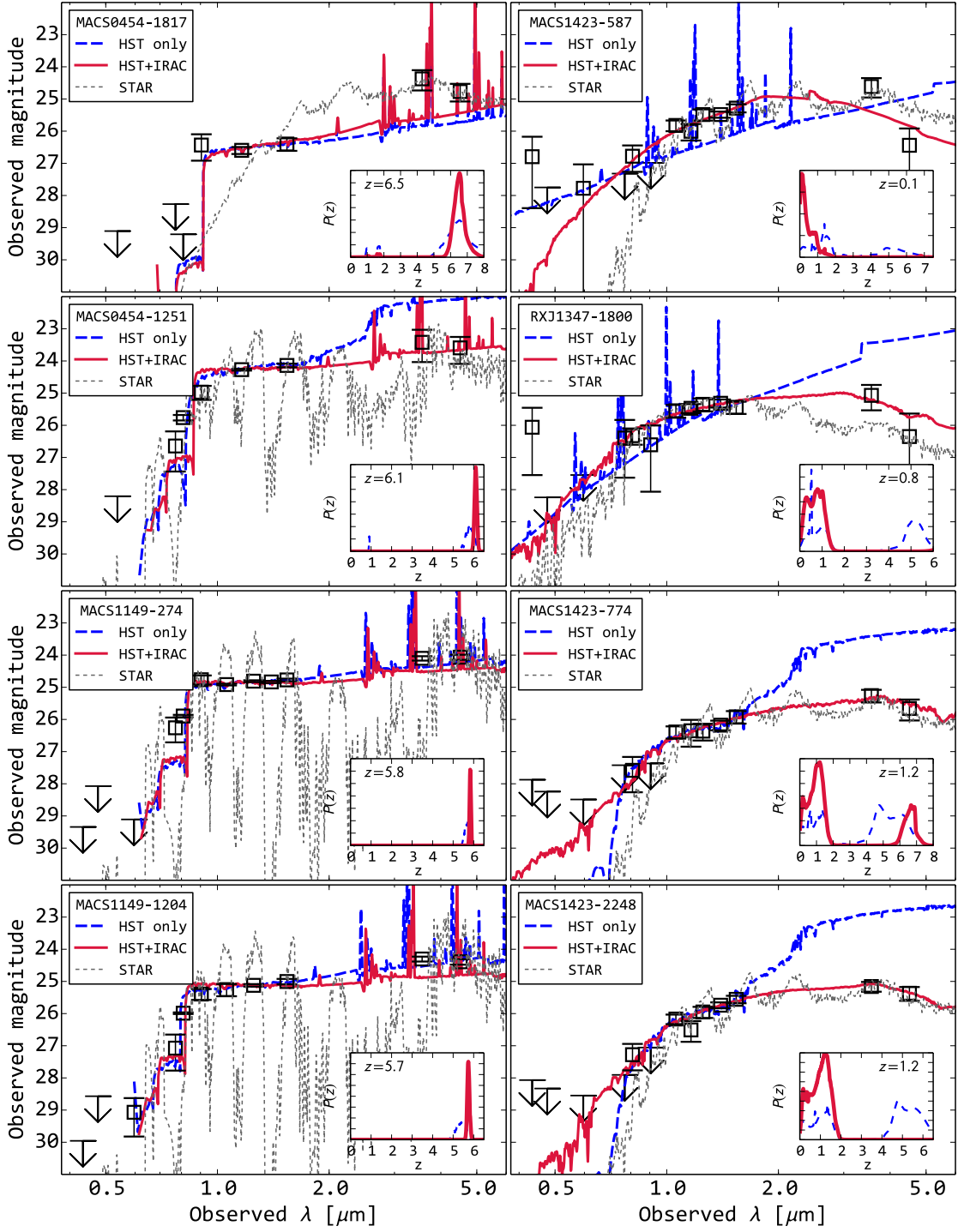}
\vspace{-0.5in}
\caption{Same as Figure \ref{fig:sedfits_all}, for the remaining eight IRAC-detected $z \gtrsim 6$ galaxy candidates. The two galaxy candidates with $z_{\text{best}} \sim 1$ are shown here.\label{fig:sedfits_all2}}
\end{figure*}

We list the best-fit galaxy properties in Table \ref{tab:sed_results} and show the best-fit templates and the photometric redshift probability density function $P(z)$ (while allowing redshift to float) in Figures \ref{fig:sedfits_all} and \ref{fig:sedfits_all2}. For each galaxy candidate, we estimate the \emph{statistical} uncertainties of stellar population properties using Monte Carlo simulations: we perturb the photometry within the errors (assuming Gaussian flux errors), re-fit with the same set of galaxy templates, and collected the distributions of each best-fit property. We only perturb the fluxes where $S/N \geq 1$. For upper limits, we do not perturb the fluxes in our simulations. The systematic errors related to assumptions in initial mass function, galaxy metallicity, and the functional form of star formation history are not represented by the error bars. We show the distributions from Monte Carlo simulations for stellar mass, star formation rate (SFR), and stellar population age in Appendix \ref{appendix_mcdist}. From these distributions, we derive the confidence intervals that bracket $68$\% of the total probability in Table \ref{tab:sed_results}. The error bars do \emph{not} include uncertainties in $\mu_{\text{best}}$.

\begin{figure}[ht]
\centering
\vspace{-1.7in}
\includegraphics[width=\columnwidth,trim=50 0 50 0, clip]{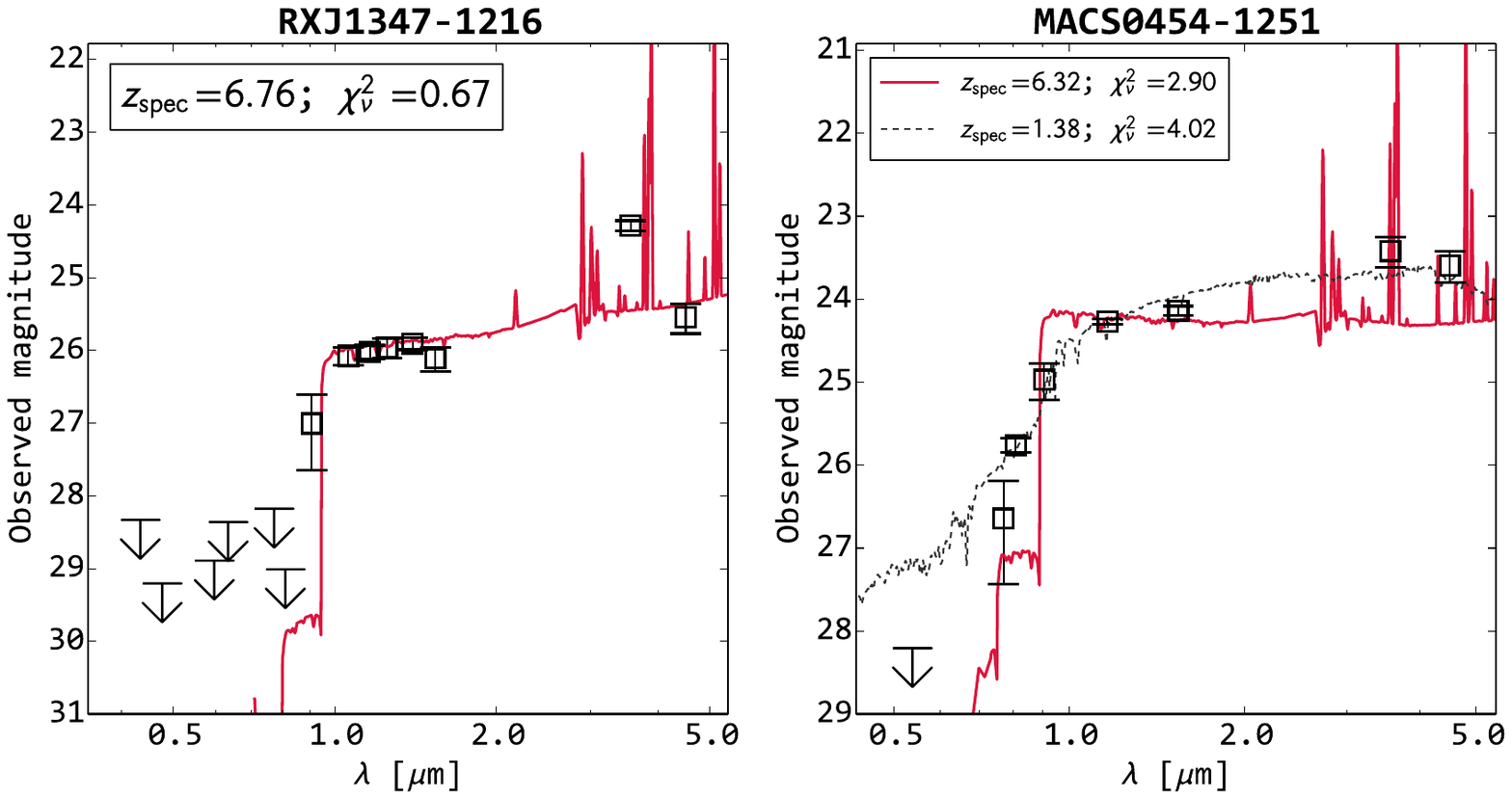}
\vspace{-1.7in}
\caption{Best-fit galaxy templates for RXJ1347-1216 and MACS0454-1251 when their redshifts are held at the Ly$\alpha$ redshifts ($z=6.76$ and $z=6.32$, respectively). For MACS0454-1251, we also show the best-fit template if the DEIMOS line detection is [OII] at $z=1.39$ instead of Ly$\alpha$. The best-fit template at $z=1.39$ has poorer fits (higher $\chi^2_{\nu}$) than the best-fit template at $z=6.32$, which supports our interpretation of the detection emission line as Ly$\alpha$. \label{fig:sedfits_wspecz}}
\end{figure}

We also show the best-fit galaxy properties for RXJ1347-1216 and MACS0454-1251, the two galaxies that we have line detections from DEIMOS data (see Section \ref{sec:spectroscopy}), when we fix their redshifts at their Ly$\alpha$ redshifts in Figure \ref{fig:sedfits_wspecz} (assuming both lines are Ly$\alpha$). For RXJ1347-1216, its photometric redshift is already sharply peaked at $z=6.7$, so the best-fit template and physical properties do not change after fixing its redshift at the Ly$\alpha$ redshift. On the other hand, MACS0454-1251 has slightly different best-fit photometric redshift and Ly$\alpha$ redshift, and we also list its best-fit properties at $z_{Ly\alpha} = 6.32$ in Table \ref{tab:sed_results}. We also show the best-fit template at $z=1.39$ in Figure \ref{fig:sedfits_wspecz} if the detected emission line is [OII] instead of Ly$\alpha$ and see that the $z=1.39$ solution has a higher $\chi^2_{\nu}$ ($\chi_{\nu}^2=4.02$) than the $z=6.32$ solution ($\chi^2_{\nu}=2.90$). The likelihood ratio of these two fits, calculated using the total $\chi^2$ as $e^{-\Delta\chi^2}$, suggests that the $z=6.32$ model is $\sim 8000$ times more likely than the $z=1.36$ model.

The best-fit stellar masses in our sample range from $2 \times 10^8\,M_\odot$ to $2.9 \times 10^9\,M_\odot$ after correcting for magnification by the foreground clusters and excluding the $z \sim 1$ interlopers. The stellar masses inferred from SED fitting have smaller statistical errors when \HST fluxes are combined with IRAC fluxes because of the constraints on rest-frame optical emission from IRAC. We show the range of stellar mass from our Monte Carlo simulations in the Appendix (Figure \ref{fig:mc_mass}), and we find that including IRAC fluxes tighten up the possible range of stellar mass for every object. We also see that the range of stellar mass spanned by our IRAC-detected sample are not necessarily at the high-mass end of the observed  (\citealt{Gonzalez:2011dn}) and simulated (\citealt{Katsianis:2015fc}) stellar mass functions at $z \gtrsim 6$. In fact, the IRAC $[3.6]-[4.5]$ color for several of our galaxy candidates (e.g., RXJ1347-1216) suggest extremely young stellar population ages ($\sim 10$Myr) and large equivalent widths from nebular emission lines. For these sources, stellar continuum emission might not dominate the observed IRAC fluxes, hence their true stellar masses depend sensitively on the equivalent widths of nebular emission lines. This demonstrates the combined power of strong gravitational lensing and deep IRAC images that allows one to measure the stellar mass of $z \gtrsim 6$ galaxies further down the stellar mass function.


On the other hand, the SFRs and stellar population ages are not necessarily well constrained by SED fitting even when IRAC fluxes are included (see Figures \ref{fig:mc_sfr} and \ref{fig:mc_age} in the Appendix). The SFRs of high-$z$ galaxies are often calculated from their rest-frame UV fluxes (after correcting for dust attenuation), and these are often the only constraints available from observations. However, the UV-derived SFR depends critically on the amount of dust attenuation inside each galaxy, and the effect of dust on the rest-frame UV color is degenerate with the effect of stellar population age. Furthermore, UV-derived SFRs probe the star formation activity over the past $\sim 100$ Myr, so it could underestimate the \emph{instantaneous} SFR if the stellar population is younger than $\sim 100\mathrm{Myr}$; for these systems, nebular emission line fluxes (e.g., H$\alpha$ or [OII]) are better proxies for SFRs (\citealt{KennicuttJr:1998vu}). We consider SFRs and stellar population ages more poorly constrained compared with stellar mass, and we will discuss the degeneracies in SED fitting in Section \ref{subsec:sedfit_error}. 

IRAC fluxes also reveal four $z \sim 1$ interlopers from our sample --- MACS1423-587, MACS1423-774, MACS1423-2248, and RXJ1347-1800 --- as shown in the three bottom-right panels in Figure \ref{fig:sedfits_all2}. All four sources have significant integrated probabilities at $z \geq 6$ when only \HST photometry is used in the fitting, but the addition of their IRAC fluxes pushes their photometric redshifts down to $z \sim 1$, suggesting that the observed breaks between F850LP and F105W are more likely the rest-frame $4000$\AA\ break instead of the Ly$\alpha$ break. This demonstrates the value of IRAC detections in discriminating between genuine $z \gtrsim 6$ galaxies and lower-$z$ interlopers. 

In Figure \ref{fig:sedfits_all} and \ref{fig:sedfits_all2} we also show the best fit stellar templates from \emph{Le Phare}. Most of the sources are better fit by galaxy templates than by stellar templates, although there is only one case, RXJ1347-1800, where both templates provide similarly good fits ($\chi_{\nu}^2=0.57$ for galaxy templates and $\chi_{\nu}^2=0.62$ for stellar templates). For all the galaxy candidates, their best-fit stellar templates are either brown dwarfs or low-mass stars from \cite{Chabrier:2000ex}. We also check if our sample contains X-ray detected sources that could have significant contributions from AGNs and we do not find evidence that any of our galaxy candidates have AGNs. However, low-level AGN activities are still possible, and the stellar population properties we infer from SED modeling will depend on how much (if any) AGN contribution there is to their broadband fluxes.


\subsection{Modeling Biases and Uncertainties}\label{subsec:sedfit_error}

\begin{figure}[t]
\vspace{-0.5in}
\includegraphics[width=\columnwidth]{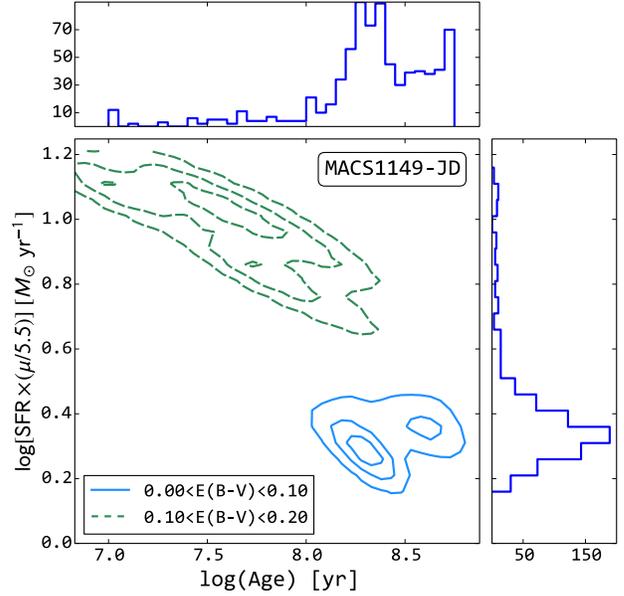}
\vspace{-0.5in}
\figcaption{Distribution of galaxy age v.s. star formation rate for MACS1149-JD ($z_{\text{phot}}=9.3$) from our Monte Carlo simulations. We plot the estimated number density contours for Monte Carlo realizations with different ranges of $E(B-V)$ values in the central panel to show the correlation between galaxy age, dust attenuation, and star formation rate. We also show the marginalized histograms of star formation rate on the right, and we show the marginalized histograms of galaxy age on the top. The star formation rate is the \emph{intrinsic} value assuming a magnification factor of $5.5$ for MACS1149-JD. \label{fig:macs1149JD_fitcor}}
\end{figure}


The biases and uncertainties of SED fitting are well documented in the literature (e.g., \citealt{Papovich:2001he,Lee:2009gj}), and most sources of systematic errors come from model assumptions. In general, stellar mass has the smallest systematic errors ($\sim 0.3$ dex), but uncertainties in galaxy star formation history can lead to large biases in galaxy age and star formation rate (\citealt{Lee:2009gj}). In additional to the usual culprits of systematic errors (e.g., star formation history, initial mass function), another important source of systematic error is the nebular emission line ratios. We find that nebular emission line contributions to broadband fluxes are important for a subset of our sample, but direct observational constraints of rest-frame optical nebular emission line ratios of $z > 6$ galaxies will not arrive until the launch of \emph{JWST}. Uncertainties in the amount of dust attenuation also complicates the interpretation of the best-fit parameters, as we demonstrate below.


We use MACS1149-JD to show how uncertainties in dust attenuation can lead to uncertainties in star formation rate in Figure \ref{fig:macs1149JD_fitcor}. We estimate the number density contours from our Monte Carlo simulations (with 1000 realizations) when IRAC fluxes are included in the modeling. We show the number density contour of each $E(B-V)$ value in the central panel and the marginalized histograms for galaxy age (on top) and star formation rate (on the right). The star formation rate histogram shows a single peak at $1.9\,\sfrunit$ assuming a magnification factor of $5.5$, but there is a long tail to higher star formation rates that extends one order of magnitude. The long tail in star formation rate corresponds to higher dust attenuation templates ($E(B-V)>0.1$; green dashed contours in the central panel) as opposed to the peak of the histogram ($E(B-V)<0.1$; solid blue contours in the central panel). We note that because of the high photometric redshift of MACS1149-JD ($z_{\mathrm{phot}}=9.3$), its UV continuum between rest-frame $1250$ and $2600$ \AA\ is not well sampled by \HST and IRAC photometry; the addition of $K$-band photometry should help constrain the amount of dust attenuation inside this galaxy.

\begin{figure}
\includegraphics[width=\columnwidth,trim=0 170 0 200,clip]{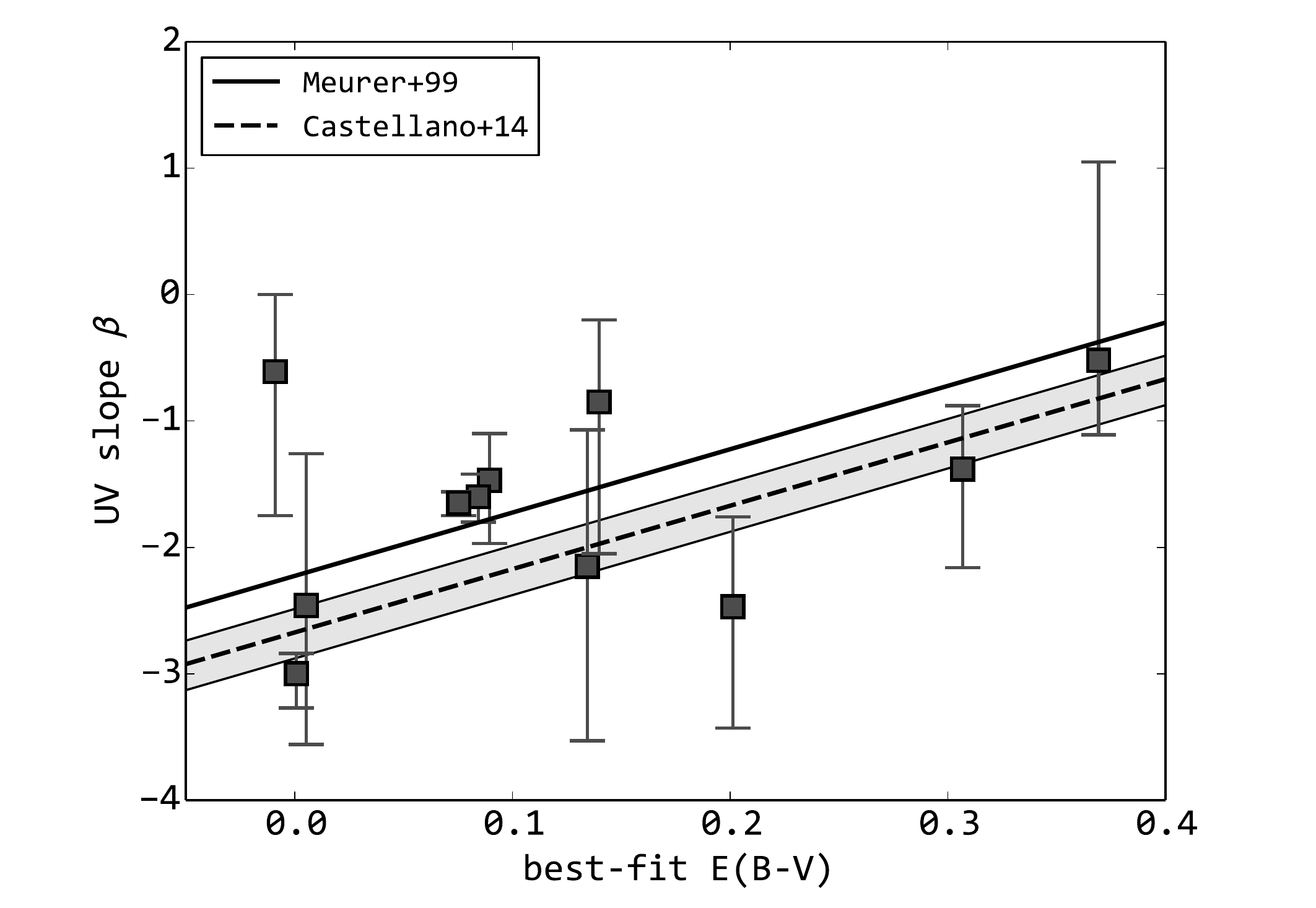}
\figcaption{Best-fit $E(B-V)$ v.s. $\beta$ for the IRAC-detected $6 \lesssim z \lesssim 10$ sample. We also show the expected $\beta$ from a given $E(B-V)$ value assuming a dust attenuation law from \cite{Calzetti:2000iy} and the empirical relations between $A_{1600}$ to $\beta$ from \cite{Meurer:1999jm} (for solar metallicity) and from \cite{Castellano:2014db} (for sub-solar metallicity). We randomly shift the best-fit $E(B-V)$ for each galaxy candidate around its best-fit value by no more than $0.01$ for clarify. As a whole sample, the measured $\beta$ values are consistent with the expected $\beta$ values from $E(B-V)$, although the scatter is large and the dust attenuation for each galaxy candidate is poorly constrained.\label{fig:ebmv_beta}}
\vspace{0.2in}
\end{figure}

As a model-independent check on the inferred dust attenuation, we measure the rest-frame UV slope $\beta$ for each galaxy candidate and list the values in Table \ref{tab:sed_results}. We can only measure $\beta$ when a galaxy candidate has at least two filters sampling the UV continuum (between rest-frame $1250$ and $2600$\AA); therefore, we do not measure $\beta$ for MACS1149-JD (which only has F160W that samples UV continuum) and for MACS1423-587, RXJ1347-1800, MACS1423-774, and MACS1423-2248 (which have photometric redshifts $\sim 1$). We measure $\beta$ by using a power-law spectrum $f_\lambda \propto \lambda^{\beta}$, convolving the spectrum with the filter curves that sample the UV continuum, and finding the $\beta$ that best matches the observed fluxes. The uncertainties are quantified in bootstrap Monte Carlo simulations, and we show the $E(B-V)$ values from SED fitting v.s. $\beta$ in Figure \ref{fig:ebmv_beta}.

In Figure \ref{fig:ebmv_beta}, we also show the expected values of $\beta$ given values of $E(B-V)$ using two different empirical calibrations. To calculate the expected $\beta$, we use the \cite{Calzetti:2000iy} dust attenuation law to calculate the amount of dust attenuation at rest-frame 1600\AA\ from $E(B-V)$: $A_{1600}=9.97 \times E(B-V)$. Then we use the relation between $A_{1600}$ and $\beta$ from \cite{Meurer:1999jm} (for solar metallicity; $A_{1600}=4.43+1.99\beta$) and \cite{Castellano:2014db} (for sub-solar metallicity; $A_{1600}=5.32^{+0.41}_{-0.37}+1.99\beta$) to calculate the expected $\beta$. We show the \cite{Meurer:1999jm} relation as a solid line and the \cite{Castellano:2014db} relation as a dashed line in Figure \ref{fig:ebmv_beta}. The scatter of the measured $\beta$ of our sample is larger than the difference between the two empirical calibrations, but the distribution is roughly consistent with both calibrations. The agreement means that the $E(B-V)$ values derived from SED fitting is not strongly biased on average, but the large scatter also suggests that for each individual galaxy, the dust attenuation is still poorly constrained.

\section{IRAC Colors and Strong Nebular Emission Lines}\label{sec:irac_colors}




Recent works suggest that at least in a subset of high-$z$ galaxies, strong nebular emission lines (most notably H$\alpha$, H$\beta$, [OIII] $\lambda 5007$, and [OII] $\lambda 3727$) with rest-frame equivalent widths $\sim 200$ \AA\ or higher contribute significantly to their IRAC fluxes (e.g., \citealt{Shim:2011cw,Schaerer:2009eq,DeBarros:2014fa,Smit:2014cg}). The galaxies with extreme nebular emission line strengths are most likely starbursts younger than $100$ Myr; such galaxies are also being found in increasing numbers at $z \sim 2$--$3$ (e.g., \citealt{vanderWel:2011ch,Atek:2011ka}). If a large number of such galaxies exist at $z \gtrsim 6$, strong nebular emission lines in the rest-frame UV/optical wavelengths need to be included in the stellar population modeling.

Within certain redshift ranges, unusual IRAC $[3.6]-[4.5]$ colors can be tell-tale signs of strong nebular emission lines. \cite{Shim:2011cw} identified 47 galaxies at $z \sim 4$ that have bluer $[3.6]-[4.5]$ colors than those expected from stellar continuum alone, and they categorized these galaxies as H$\alpha$ emitters because the SED bumps at $3.6\mu$m are likely due to strong H$\alpha$ emission. More recently, \cite{Smit:2014cg} and \cite{Smit:2015jc} presented $z \sim 6.6$--$7.0$ galaxy candidates with unusually blue $[3.6]-[4.5]$ colors as evidence for strong contributions from [OIII] and H$\beta$ to the $3.6\mu$m fluxes. The colors of these peculiar objects usually can only be reproduced with model SEDs that include strong nebular emission lines.


\begin{figure}[t]
\vspace{-0.9in}
\includegraphics[width=\columnwidth,trim=0 120 0 0,clip]{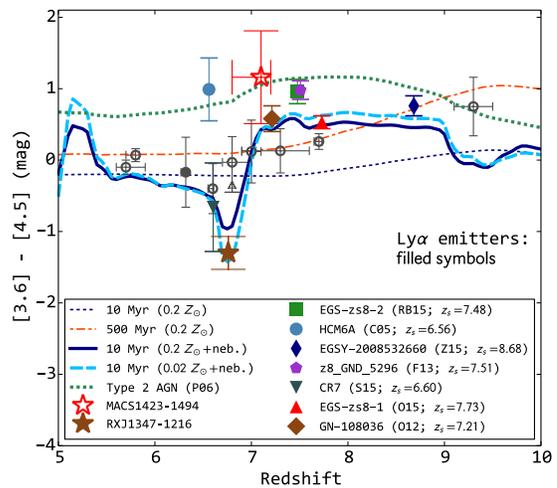}
\figcaption{IRAC $[3.6]-[4.5]$ color as a function of redshift calculated from BC03 model spectra. We show the $[3.6]-[4.5]$ colors for $0.2\ Z_{\odot}$ stellar population models that are 500 Myr old \emph{without} nebular emission lines (thin dot-dashed curve), 10 Myr old \emph{without} emission lines (thin dotted curve), and a 10 Myr old model \emph{with} nebular emission lines (thick solid curve). In our implementation, the dust-free 10 Myr model with nebular emission lines has H$\alpha$, H$\beta$, and [OIII]$\lambda 5007$ equivalent widths of $1087$\AA, $182$\AA, and $868$\AA, respectively; for the $10$ Myr model, the equivalent widths change by $< 3$\% within the range of $e$-folding time $\tau$ that we adopt. In addition to the fiducial $0.2\ Z_{\odot}$ models, we also show the expected colors of a $0.02\ Z_{\odot}$, 10 Myr old model with nebular emission lines for comparison (thick dashed curve); the more metal-poor model seems to reproduce the colors of RXJ1347-1216. The measured $[3.6]-[4.5]$ colors of two sources in our sample that likely have strong emission lines are shown in stars, and the rest of the sample is shown in black circles. We show the known Ly$\alpha$ emitters (LAEs) in \emph{filled symbols}, including the two LAEs reported in this work (see Section \ref{sec:spectroscopy}). We also show other published $z > 6$ LAEs with measured $[3.6]-[4.5]$ colors --- z8\_GND\_5296 (\citealt[F13]{Finkelstein:2013fx}), HCM6A (\citealt[C05]{Chary:2005be}), GN-108036 (\citealt[O12]{Ono:2012kv}), CR7 (\citealt[S15]{Sobral:2015fq}), EGS-zs8-1 (\citealt[O15]{Oesch:2015ky}), EGS-zs8-2 (\citealt[RB15]{RobertsBorsani:2015um}), and EGSY-2008532660 (\citealt[Z15]{Zitrin:2015to}) --- in \emph{filled} symbols. The $[3.6]-[4.5]$ color of z8\_GND\_5296, HCM6A, EGS-zs-8-2, and MACS1423-1494 are hard to reproduce by the stellar population models that we adopt, but they come close to the expected colors of a Type 2 (obscured) AGN template from \cite[P06]{Polletta:2006jt}. \label{fig:irac_colors}}
\end{figure}

In Figure \ref{fig:irac_colors}, we compare the IRAC $[3.6]-[4.5]$ colors of our sample with a range of model predictions. We use our fiducial SED model (BC03) to generate the redshift evolution of $[3.6]-[4.5]$ color at 500 Myr old \emph{without} nebular emission lines (thin dot-dashed curve, roughly the age of the universe at $z=9.5$), 10 Myr old \emph{without} nebular emission lines (thin dotted curve), and 10 Myr old \emph{with} nebular emission lines (thick solid curve). The 10 Myr old model with nebular emission lines have equivalent widths $1087$\AA, $182$\AA, and $868$\AA\ for H$\alpha$, H$\beta$, and [OIII]$\lambda\lambda 4959,5007$, respectively. Here we assume the star formation $e$-folding time scale $\tau$ to be 100 Myr, but the $[3.6]-[4.5]$ color does not change significantly when different values of $\tau$ are used.


The most prominent feature in Figure \ref{fig:irac_colors} is the ``dip'' in $[3.6]-[4.5]$ for a 10 Myr old starburst with nebular emission lines at $z \sim 6.8$ due to the contributions from [OIII] and H$\beta$ --- the same feature that \cite{Smit:2014cg} utilized to identify strong nebular emission line objects within $6.6 \lesssim z \lesssim 7$. In our sample, only RXJ1347-1216 has a photometric redshift $\sim 6.8$ and a very blue $[3.6]-[4.5]$ color. This source has a best-fit age of $10$ Myr, the youngest age included in our templates. Extremely young stellar populations are expected to generate a large number of ionizing photons, so if these sources are indeed $\sim 10$ Myr old starbursts, they might also have high Ly$\alpha$ luminosities around star forming regions. We already successfully identified one of the three sources (RXJ1347-1216) as a $z = 6.76$ Ly$\alpha$ emitter (LAE; see Section \ref{sec:spectroscopy}); we do not identify other sources at $z \sim 6.8$ with blue $[3.6]-[4.5]$ color that could also be strong line emitters in our sample.

In Figure \ref{fig:irac_colors} we also show the redshift evolution of $[3.6]-[4.5]$ color for a 10 Myr old, $0.02Z_{\odot}$ model (thick dashed curve), and it predicts a bluer $[3.6]-[4.5]$ color at $z \sim 6.8$ (as blue as $\sim -1.4$ mag) than the 10 Myr old, $0.2Z_{\odot}$ model. The IRAC colors of the $0.02Z_{\odot}$ model at $z \sim 6.8$ show better agreements with the three sources mentioned above than the $0.2Z_{\odot}$ model, which suggests that these sources might have lower metallicities than our fiducial model. We note that the nebular emission line properties of individual galaxies are highly uncertain (and are sensitive to metallicity), so any constraint on metallicity is preliminary.

Our fiducial galaxy SED models also predict that young starbursts with strong nebular emission lines should have red $[3.6]-[4.5]$ colors at $7.0 \lesssim z \lesssim 7.5$. In this redshift range, [OII] and [OIII] move into IRAC ch1 and ch2, respectively, and the combined [OIII]$+$H$\beta$ line flux is expected to be $\sim 4$ times higher than [O II] in our implementation (\citealt{Anders:2003ci}); the expected $[3.6]-[4.5]$ color reaches $\sim 0.5$ mag within $7.0 \lesssim z \lesssim 7.5$. MACS1423-1494 (\photz$=7.3$) has a photometric redshift and measured $[3.6]-[4.5]$ color that are close to the model prediction, although its red $[3.6]-[4.5]$ color is hard to reconcile even with the 10 Myr old galaxy model. Based on its photometric redshift and unusually red $[3.6]-[4.5]$ color (it is also best-fit by a 10 Myr old galaxy SED), we identify this source as another prime candidate for Ly$\alpha$ emission. We note that Ly$\alpha$ photons are subject to complicated radiation transfer effects both inside and outside of galaxies, so it is far from guaranteed that these sources will have detectable Ly$\alpha$ emission. But they are likely LAE candidates (compared with other high-$z$ galaxies) based on their photometric redshifts and IRAC colors.

We also compare our galaxy model-predicted IRAC colors with other $z \gtrsim 6.5$ LAEs with published IRAC colors in Figure \ref{fig:irac_colors}. The other LAEs include HCM6A from \cite{Hu:2002hb} ($z_s=6.56$, where $z_s$ is the spectroscopic redshift determined by Ly$\alpha$ emission), CR7 from \cite{Sobral:2015fq} ($z_s=6.60$), GN-108036 from \cite{Ono:2012kv} ($z_s=7.21$), EGS-zs8-2 from \cite{RobertsBorsani:2015um} ($z_s=7.48$), z8\_GND\_5296 from \cite{Finkelstein:2013fx} ($z_s=7.51$), EGS-zs8-1 from \cite{Oesch:2015ky} ($z_s=7.73$), and EGSY-2008532660 from \cite{Zitrin:2015to} ($z_s=8.68$). All of these LAEs have IRAC colors that strongly suggest high nebular emission line equivalent widths (most likely [OIII] and H$\beta$ at this redshift range), because they lie along the curve traced by a dust-free, $0.2\ Z_\odot$, 10 Myr stellar population. For example, \cite{Finkelstein:2013fx} argued that the red IRAC color of z8\_GND\_5296 is due to the galaxy's strong [OIII]$+$H$\beta$ emission lines in IRAC ch2, and they inferred the [OIII] $\lambda 5007$ equivalent width to be 560--640 \AA\ from photometry. The IRAC colors of these $z \gtrsim 6.5$ LAEs corroborate the recent findings that many galaxies detected at $z \gtrsim 6$ likely have high nebular emission line equivalent widths.

Two notable cases among the group of LAEs in Figure \ref{fig:irac_colors} are MACS1423-1494 and HCM6A\footnote{HCM6A was first reported by \cite{Hu:2002hb}, and later \cite{Chary:2005be} published its IRAC fluxes.}. HCM6A was found in the vicinity of a massive galaxy cluster Abell 370\footnote{Abell 370 is one of the Hubble Frontier Fields cluster.} and has a measured $[3.6]-[4.5]$ color of $1.0 \pm 0.4$ mag, significantly redder than the $[3.6]-[4.5]$ color predicted by a 10 Myr stellar population model at its redshift ($z_s=6.56$). The red $[3.6]-[4.5]$ color suggests a very high H$\alpha$/([OIII]$+$H$\beta$) ratio, which is unexpected (but not impossible) for a young, low-metallicity stellar population. In order to explore other possibilities to explain the red $[3.6]-[4.5]$ colors of both LAEs, we plot the predicted $[3.6]-[4.5]$ colors of a Type 2 obscured AGN template from \cite{Polletta:2007ha}. This obscured AGN template includes a dust attenuation of $A_V = 4$ mag that fits the obscured AGN SW 104409 ($z=2.54$; \citealt{Polletta:2006jt}), and its color trajectory in redshift is shown as a thick dotted curve in Figure \ref{fig:irac_colors}. Interestingly, the predicted $[3.6]-[4.5]$ colors of an obscured AGN agrees quite well with the colors of both MACS1423-1494 and HCM6A, and z8\_GND\_5296 and EGS-zs8-2 also have marginally consistent IRAC colors with this obscured AGN template. If these sources indeed harbor obscured AGNs (like SW 104409), the red $[3.6]-[4.5]$ colors will be primarily due to large dust attenuation in the rest-frame optical, while the blue rest-frame UV colors come from the scattered light of the central QSO emission. Obscured AGN is an intriguing possibility to consider for these sources, although so far no direct evidence exists that any of these sources have significant flux contributions from an obscured AGN.

To sum up, we identify three sources in our sample at $z \sim 6.7$ and $z \sim 7.3$ as likely young starbursts with very strong nebular emission lines based on their IRAC colors. We detect Ly$\alpha$ emission in one of them, RXJ1347-1216, during our recent DEIMOS observations, and we plan to follow up all the other three sources for their potential Ly$\alpha$ emission. 






\section{Summary}\label{sec:summary}

In this work, we present the constraints on the $6 \lesssim z \lesssim 10$, IRAC-detected galaxy candidates behind eight strong-lensing galaxy clusters from SURFS UP. Six of the clusters are in the CLASH sample, and two are in the Hubble Frontier Fields sample. We summarize our findings as follows:

\begin{itemize}

\item We find a total of 17 galaxy candidates using the Lyman break color selection that have $S/N \geq 3$ in at least one IRAC channel. The photometric redshifts in our sample range from $5.7$ to $9.3$, and we identify four galaxy candidates (MACS1423-587, RXJ1347-1800, MACS1423-774, and MACS1423-2248) as likely $z \sim 1$ interlopers after including their IRAC fluxes in the SED modeling. We find the largest number (6) of IRAC-detected galaxy candidates in MACS1423.

\item From our Keck spectroscopic observations, we identify one secure Ly$\alpha$ emitter at $z = 6.76$ (RXJ1347-1216) and one likely Ly$\alpha$ emitter at $z = 6.32$ (MACS0454-1251). The line equivalent widths, assuming they are both Ly$\alpha$, are $26 \pm 4$\AA\ (RXJ1347-1216) and $6.8 \pm 1.7$\AA\ (MACS0454-1251, averaged over two nights). We infer lower limits of their star formation rates from their Ly$\alpha$ line fluxes and find them to be consistent with the star formation rates from SED fitting.

\item We infer the physical properties of our sample galaxies using \cite{Bruzual:2003ck} galaxy templates and add nebular emission lines to the templates. Under our SED modeling assumptions ($0.2\ Z_{\odot}$, Chabrier IMF, exponentially decaying star formation history, and nebular line emission), the stellar masses of our sample range from $0.2$--$2.9\times 10^{9}\ M_{\odot}$ (excluding the three likely $z \sim 1$ interlopers) when we use the best available magnification factors for each galaxy candidate. The magnification-corrected rest-frame 1600 \AA\ absolute magnitude ($M_{1600}$; see Table \ref{tab:sed_results}) of our sample ranges from $-21.2$ to $-18.9$ mag. The range of intrinsic UV luminosity probed here is slightly fainter than the knee of UV luminosity functions at $6 \lesssim z \lesssim 10$, which have $\mstar$ between $\sim -20.6$ and $\sim -21.6$ mag (e.g., \citealt{Bouwens:2015gm,Finkelstein:2015ir}), showing that galaxy clusters' strong lensing power allows us to start probing the more typical UV luminosities. The range of intrinsic stellar mass probed here is also close to the knee of the stellar mass functions at this redshift range (e.g., \citealt{Gonzalez:2011dn,Katsianis:2015fc}). Some galaxies in our sample are best fit by extremely young ($\sim$10 Myr old) templates and others best fit by more evolved (up to $\sim$ 700 Myr old at $z \sim 7$) templates, suggesting that the IRAC-detected sample contains both very young galaxies with strong nebular emission lines and more evolved and massive galaxies at $6 \lesssim z \lesssim 10$. 


\item From the photometric redshifts and IRAC colors, we identify two galaxy candidates that likely have strong (rest-frame optical) nebular emission lines: RXJ1347-1216 and MACS1423-1494. Both sources are best fit by the youngest (10 Myr old) galaxy templates included in our modeling and are prime targets for spectroscopic observations. We already identified one of them (RXJ1347-1216) as a Ly$\alpha$ emitter, and we will target the other two in our future spectroscopic observations. Other galaxies in the sample lie in the part of the redshift--IRAC color space that makes it hard to infer their nebular emission line strengths; namely, they are within the redshift range that both IRAC bands could have contribution from strong nebular emission lines such as [OIII], H$\beta$, and H$\alpha$, and their IRAC colors may not be very different from those of pure stellar continuum.

\end{itemize}

The IRAC fluxes provide important information about the galaxies at $6 \lesssim z \lesssim 10$, because it is the only probe of their rest-frame optical emission that we have at the moment. The IRAC-detected galaxies may not be representative of the entire galaxy population at $z \gtrsim 6$, but their IRAC colors do provide a more effective way to select spectroscopic targets for redshift confirmation. IRAC fluxes and meaningful upper limits can also distinguish some lower-redshift galaxies from high-$z$ dropouts and are important for constructing clean $z \gtrsim 6$ galaxy samples.
\\
\\
We would like to thank the anonymous referee for constructive suggestions that make this work better. We also thank Harry Ferguson, Samuel Schmidt, Chris Fassnacht, Dennis Zaritsky, and Hendrik Hildebrandt for useful discussions and comments on the manuscript. Observations were carried out using Spitzer Space Telescope, which is operated by the Jet Propulsion Laboratory, California Institute of Technology under a contract with NASA. Also based on observations made with the NASA/ESA Hubble Space Telescope, obtained at the Space Telescope Science Institute, which is operated by the Association of Universities for Research in Astronomy, Inc., under NASA contract NAS5-26555 and NNX08AD79G and ESO-VLT telescopes. Support for this work was provided by NASA through a Spitzer award issued by JPL/Caltech. This work was supported by NASA Headquarters under the NASA Earth and Space Science Fellowship Program - Grant ASTRO14F- 0007. We also acknowledge support from HST-AR-13235, HST-GO-13177, and special funding as part of the HST Frontier Fields program conducted by STScI. TS acknowledges support from the German Federal Ministry of Economics and Technology (BMWi) provided through DLR under project 50 OR 1308. TT acknowledges support by the Packard Fellowship. The Dark Cosmology Centre (DARK) is funded by the Danish National Research Foundation.

\appendix

\section{Distributions from Monte Carlo Simulations}\label{appendix_mcdist}
\begin{figure*}[t]
\centering
\vspace{-1in}
\includegraphics[width=\textwidth]{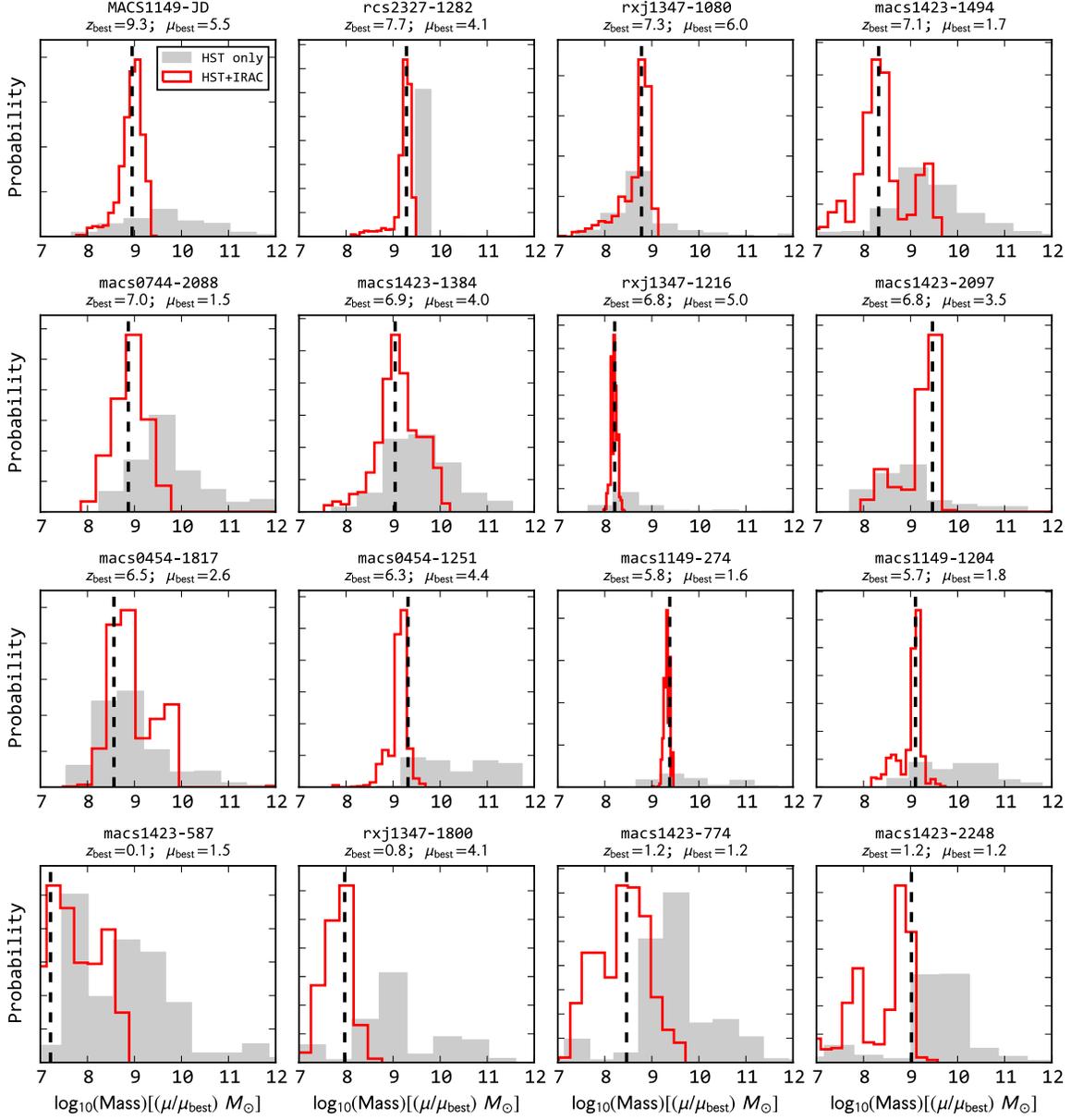}
\vspace{-1in}
\caption{Stellar mass distributions derived from Monte Carlo simulations for each IRAC-detected $z \gtrsim 6$ galaxy candidate. The stellar mass values have been scaled by our best estimates of magnification factors $\mu_{\text{best}}$. The distributions from combining \HST and IRAC photometry are shown as the unfilled red histograms; the distributions from \HST photometry only are shown as filled gray histograms. The best-fit stellar masses from Table \ref{tab:sed_results} are shown as the vertical dashed lines.\label{fig:mc_mass}}
\end{figure*}

\begin{figure*}[hp]
\centering
\vspace{-1in}
\includegraphics[width=\textwidth]{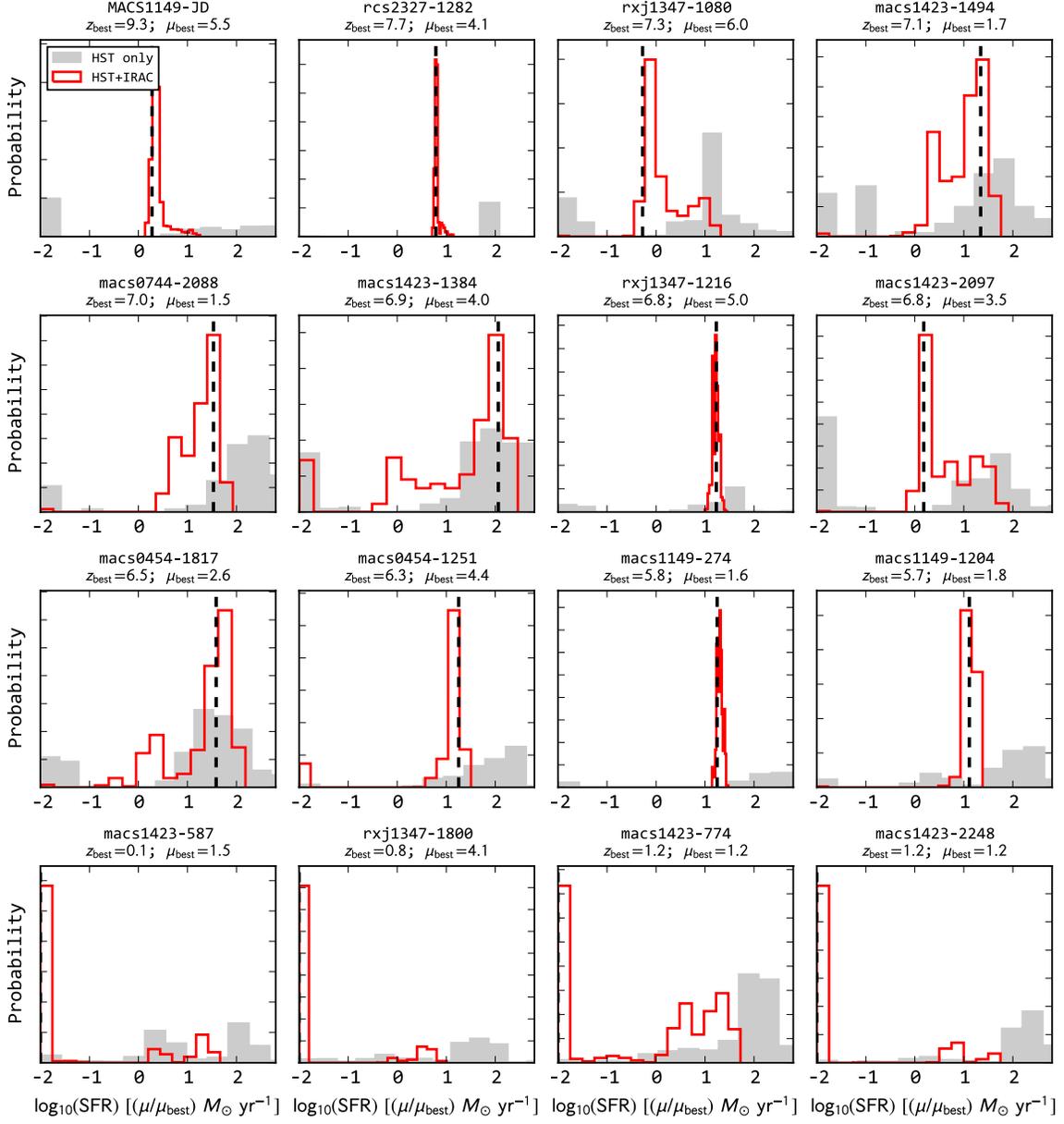}
\vspace{-1in}
\caption{Star formation rate (SFR) distributions derived from Monte Carlo simulations for each IRAC-detected $z \gtrsim 6$ galaxy candidate. The SFR values have been scaled by our best estimates of magnification factors $\mu_{\text{best}}$. All SFRs below $0.01\,M_{\odot}\,\text{yr}^{-1}$ are set to $0.01\,M_{\odot}\,\text{yr}^{-1}$ for clarity. The distributions from combining \HST and IRAC photometry are shown as the unfilled red histograms; the distributions from \HST photometry only are shown as filled gray histograms. The best-fit SFR from Table \ref{tab:sed_results} are shown as the vertical dashed lines. \label{fig:mc_sfr}}
\end{figure*}

\begin{figure*}[hp]
\centering
\vspace*{-1in}
\includegraphics[width=\textwidth]{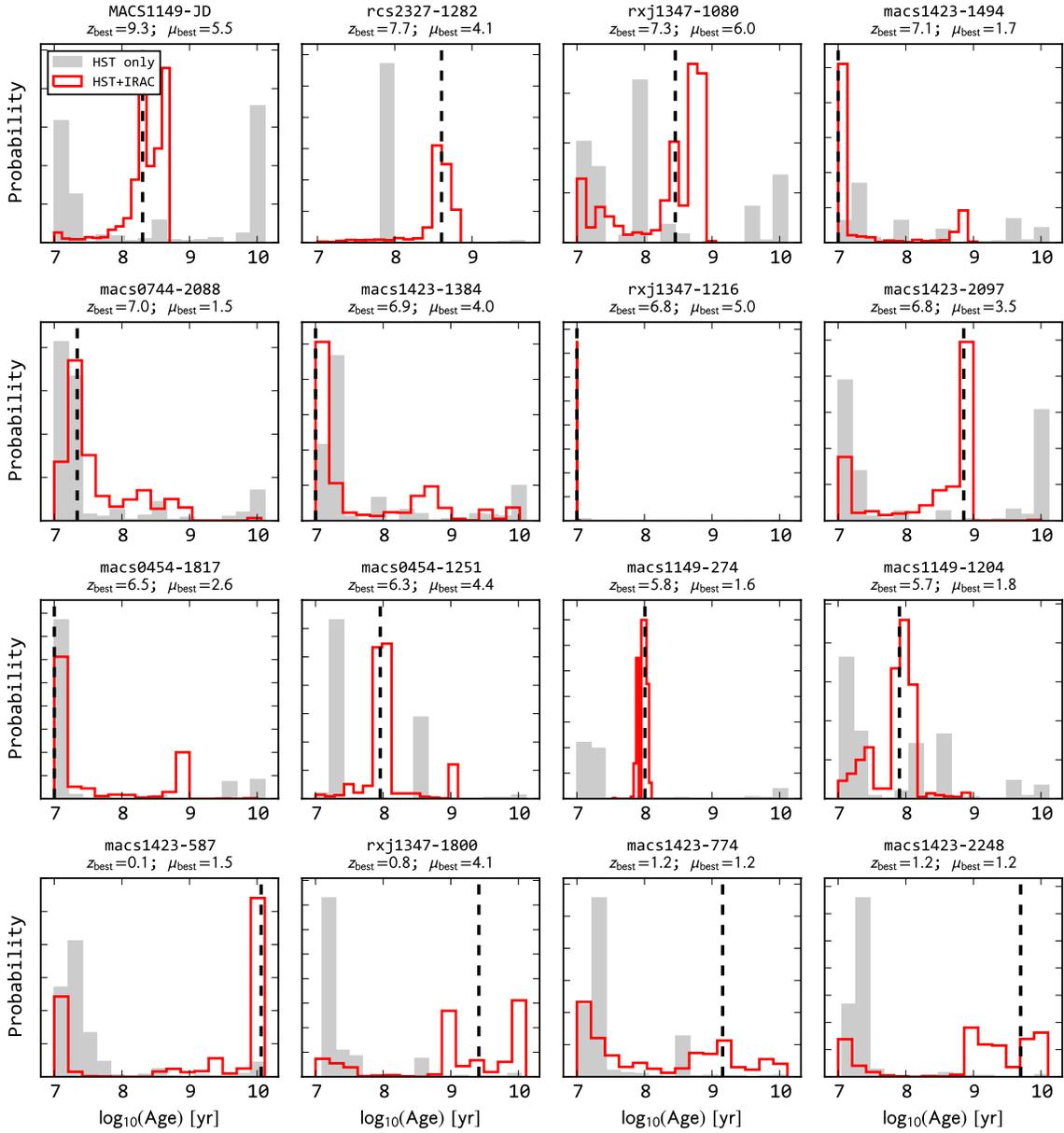}
\vspace{-1in}
\caption{Model stellar population age distributions derived from Monte Carlo simulations for each color-selected, IRAC-detected $z \gtrsim 6$ galaxy candidate. The minimum age included in the template library is $10$ Myr. The distributions from combining \HST and IRAC photometry are shown as the unfilled red histograms; the distributions from \HST photometry only are shown as filled gray histograms. The best-fit age from Table \ref{tab:sed_results} are shown as the vertical dashed lines. \label{fig:mc_age}}
\end{figure*}

Here we show the distributions of stellar mass, star formation rate, and stellar population age (assuming an exponentially declining star formation history with $e$-folding time between $0.1$ and $30$ Gyr) in Figures \ref{fig:mc_mass}, \ref{fig:mc_sfr}, and \ref{fig:mc_age}, respectively. In all panels, the distributions from using \HST photometry only are shown in gray filled histogram, while the distributions from combining \HST and \spitzer photometry are shown in red histogram.

\clearpage
\bibliographystyle{apj}
\bibliography{surfsup2_sub_arxiv2}

\end{document}